# Mechanical Properties of the Meninges: Large Language Model Assisted Systematic Review of over 25,000 Studies

January 31, 2024

Brandon P. Chelstrom[1], Maciej P. Polak[2], Dane Morgan[2], Corinne R. Henak[1,3,4]

[1]Department of Biomedical Engineering, University of Wisconsin-Madison, Madison, WI
[2]Department of Materials Science and Engineering, University of Wisconsin-Madison, Madison, WI
[3]Department of Mechanical Engineering, University of Wisconsin-Madison, Madison, WI
[4]Department of Orthopedics and Rehabilitation, University of Wisconsin-Madison, Madison, WI

*Address Correspondence to:
Corinne R. Henak, PhD
Associate Professor
3031 Mechanical Engineering Building
1513 University Ave
Madison, WI 53706
608-263-1619
chenak@wisc.edu

Word count: 5949 words

**Graphical Abstract:**

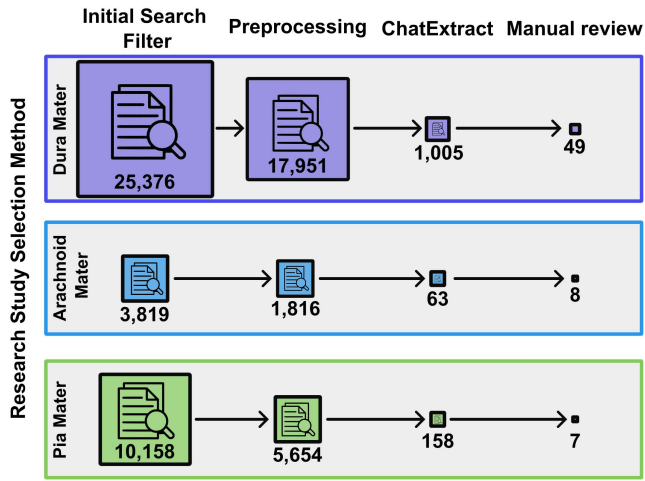

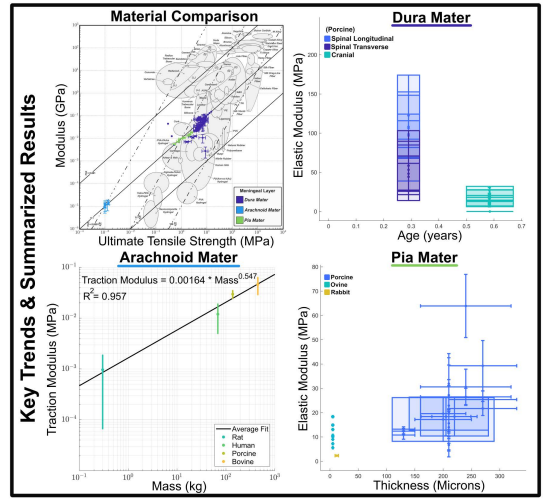

**Abstract:**

Accurate constitutive models and corresponding mechanical property values for the meninges are important for predicting mechanical damage to brain tissue due to traumatic brain injury. The meninges are often oversimplified in current finite element (FE) head models due to their complex anatomy and spatially-variant mechanical behavior. This study performed a systematic review (SR) on the mechanical properties of each individual layer of the meninges to obtain benchmark data for FE modeling and to identify gaps in the current literature. Relevant studies were filtered through three stages: a broad initial search filter, a large language model classifier, and manual verification by a human reviewer. Out of over 25,000 studies initially considered, this review ultimately included 47 studies on the *dura mater*, 8 on the *arachnoid mater*, and 7 on the *pia mater*, representing the largest and most comprehensive SR on the mechanical properties of the meninges. Each layer was found to exhibit nonlinear rate dependence that varies with species, age, location, and orientation. This study revealed that the elastic modulus of *pia mater* most often used in simplified linear elastic FE models is likely underestimated by an order of magnitude and fails to consider directional dependence. Future studies investigating the mechanical properties of the meninges should focus on a wider range of loading rates as well as age effects for the arachnoid mater and pia mater, as these features are relatively understudied and expected to affect the fidelity of FE predictions.

## 1) Introduction:

Traumatic brain injury (TBI) occurs due to head trauma leading to mechanical damage of brain tissue. The yearly fatality rate from TBI in the United States has remained relatively constant at 17 per 100,000 from 2000 to 2017 [1], while TBI hospitalization rates were considerably higher at 70.1 per 100,000 in 2017 [2], both of which disproportionally affect youth athletes [1–3], elderly populations [1,2,4], and military personal (40-60% of deployed service members [5]). Actual rates of TBI-related injuries are likely underestimated [3], as individuals with mild head trauma often do not seek treatment despite experiencing short-term neurological impairment [6,7]. Clinicians are similarly limited in their ability to treat TBI, often stuck between waiting for symptoms [8] or severe surgical options largely focused on preventing secondary damage [9]. Critical to improving both treatment and understanding of TBI are finite element (FE) models, which estimate cellular injury from organ- and tissue-level loads. Kinematic full head simulations predict that angular velocity and acceleration can lead to strain levels that damage neurons [10–14] due to the relative movement of the cortex inside the skull [15–19], leading to both surface [11,12,20,21] and subsurface [10,12] strains above proposed neuronal injury thresholds [21–23].

Accurate mechanical properties are essential to accurate FE predictions. The cortex and the skull are directly connected [24,25] via the meninges, a thin fibrous tissue composed of three layers identifiable via gross morphology: the *dura mater* (400 µm – 1400 µm thick in humans [26]), *arachnoid mater* (20 µm – 53 µm thick in humans [27]), and *pia mater* (4.96 ± 0.82 µm thick in sheep [28]). Meninges primarily contain fibroblasts [29–31], type 1 collagen fibers [32], proteoglycans [32], cerebral spinal fluid (CSF) [24,30], and blood vessels [33]. Individual layers and the connections between adjacent layers are highly heterogenous with respect to collagen content [29,32,34,35], collagen fiber alignment [26,34,36,37], cell phenotype [29,32,38], cell density [29,32,38], porosity [29,30,39], and vasculature density [25,36,37]. Following the structure, the mechanical properties of the meningeal layers exhibit heterogeneity [40,41] with respect to rate-dependence [28,42], nonlinearity [27,43], and poroviscoelasticy [44–46]. The combination of both complex anatomy and mechanical properties has often led to an over-simplification of the brain-skull interface in FE head models [20,47,48] which drastically changes the predicted distribution of cortical surface strain [19,20], and thereby affects the accuracy of the estimation of cellular injury.

Systematic reviews (SR) can often address disparities between studies by synthesizing data on material behavior and anatomical structure. A collection of recent SRs has summarized elastic tissue properties but lack quantitative analysis on more complex tissue behavior [20,49,50]. Importantly, several recent studies have highlighted the importance of meningeal mechanical properties in injury prediction. One FE study found that predictions of neuronal strain were highly dependent on mechanical behavior of the brain-skull interface [20]. A review of spinal and cranial *dura mater* mechanical properties found the most commonly used elastic modulus of 31.5 MPa was likely underestimating the true tissue modulus by a factor of two [49]. Prior SRs also highlighted that the oversimplification common in FE head models arises from the lack of dynamic loading rates (>120 mm/min), the lack of anisotropic mechanical properties, and the lack of reporting in tissue heterogeneity as important factors for accurately modeling TBI [49,50]. These prior SRs only considered elastic tissue properties [49], or considered the *arachnoid mater* and *pia mater* as a single mechanical unit [50]. Thus, there remains a need for an SR that collates complete mechanical behavior of all meningeal tissues.

One of the biggest challenges in a standard SR [51] is method used to identify all unique primary studies on a given topic. In a fully manual SR, the reviewer would most likely only consider the full text of a study if the abstract includes a predetermined set of criteria describing a primary study. Identifying abstracts of interests is often done using search filters to limit both the review's scope to topic(s) of interest and the volume of abstracts to screen [52], leading to limitations in reviews as described above. Search filters are limited by the choice search terms, which often vary considerably between SRs [52]. Attempts have been made to create field-specific, generalized search filters that increase sensitivity and specificity, but at a cost of reduced reviewer efficiency due to article screening time [53]. A simulation of the SR process using a minimum of 7 reviewers found that a median of 881 person-hours (corresponding to ~22 standard 40-hr work weeks) was required to complete an SR that identified 20 relevant articles out of 3,831 articles on average [54]. With the volume of scientific publications following an annual exponential growth rate [55], the time required to conduct a full SR is also expected to grow exponentially. Additionally, a traditional SR which only screens abstracts risks the exclusion of relevant research if it is not sufficiently emphasized in the abstract. One solution is to use automated systems, such as natural language processing, to automatically scan publications [56,57]. Fully automated solutions are less accurate than human reviewers, with one fully automatic SR using a large language model (LLM) only able to accurately identify 67% of articles found by human reviewers [58]. A potential resolution to the limited search space of a standard SR and the accuracy of a fully automatic SR is a two-stage SR, first analyzed by an LLM followed by manual evaluation by a human reviewer (LLM-assisted SR) [59].

Therefore, the objective of this study is to perform an LLM-assisted SR on the mechanical properties of the meninges, providing benchmark data for FE modeling and highlighting areas where further experimental results would be most impactful. A preliminary collection of studies (research papers) was gathered using general and broad search filters to complete an exhaustive search of the current available literature. Relevant data in research papers was identified using ChatExtract [60], an LLM-based workflow for automated mechanical data extraction from research texts. The output from ChatExtract was then manually evaluated to only include data from primary mechanical testing papers which was further manually expanded to include additional relevant features. A hybrid approach combines the high-throughput capabilities of automated LLM-based methods with human expertise to improve overall accuracy, resulting in the largest SR on the mechanical properties of the meninges. By gathering more data, the interactions between age, species, loading rate, thickness, and constitutive model parameters as well as their importance for on FE models were investigated.

## 2) Methods

### 2.1) Initial Article Selection

The majority of this study followed the most current Preferred Reporting Items for Systematic Review and Meta-Analysis (PRISMA) guidelines [51] (Fig. 1). The initial search space was constrained to studies available under the UW-Madison license published from October 1823 to May 2024 from ScienceDirect. Other databases were not considered due to limited automated access to full text articles. It is worth noting that the human reviewer focused on evaluating the LLM extracted data for primary mechanical studies as well as references to primary mechanical study. Referenced studies were obtained manually and included journals from outside ScienceDirect, effectively covering the ScienceDirect database along with studies referenced by potentially relevant papers.

The initial search filter was made intentionally broad by only including strict material property pairs in the format: "{Material}" +"AND" +" {Property}". Each individual layer of the meninges, the *dura mater, arachnoid mater,* and *pia mater*, were selected as the material terms. Mechanical property terms included both specific (e.g., elastic modulus) and general (e.g., FE parameter) parameters. The latter captured studies that included more complex constitutive models. The full parameter set for properties included specific parameters (elastic modulus, Young's modulus, shear modulus, relaxation time, strain, strain rate, and rate dependence), and general parameters (stiffness, nonlinear, temperature, linear elastic, viscoelastic, pressure, finite element parameter, and model). Following the preliminary search, duplicate articles and articles that contained less than 1000 characters, found to be abstracts, conference headings, or null entries, were excluded.

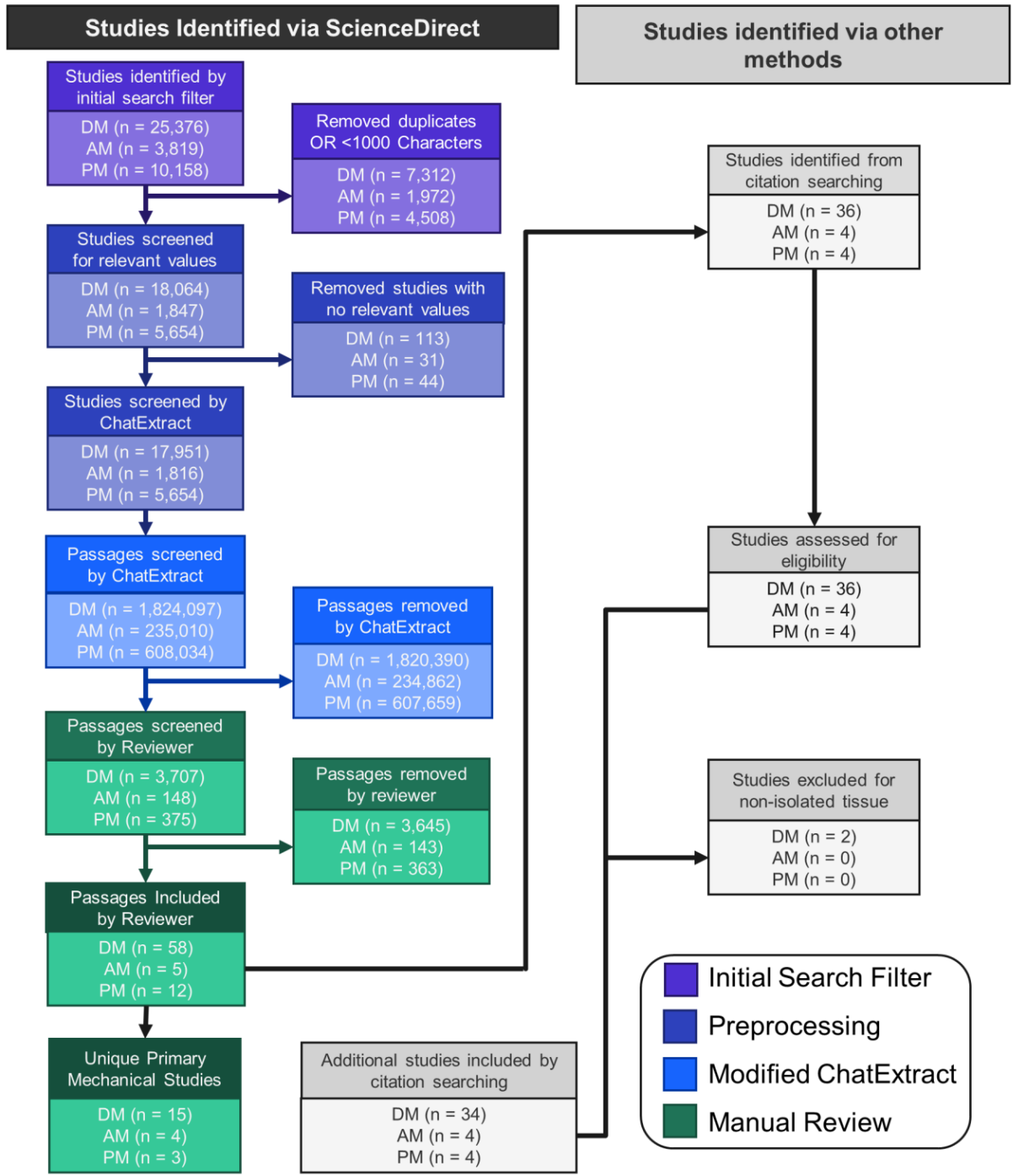

Fig.1: PRISMA flow chart of study search procedure for identifying relevant literature based on current 2020 PRIMSA reporting guidelines [51]. Purple boxes denote the initial search filer for the dura mater (DM), arachnoid mater (AM), and pia mater (PM). Dark blue boxes correspond to the LLM processing, light blue boxes denote automated processing by the ChatExtract algorithm [60], and green boxes signify studies identified by manual review. Light gray boxes correspond to studies identified via the snowball search method. The conversion from studies to passages marks the transition from the preliminary relevance stage which serves to drastically reduce the search space for the LLM.

### 2.2) Motivation of the LLM-assisted SR approach

This review used ChatExtract [60] to gather structured data on the mechanical properties of the *dura mater, arachnoid mater,* and *pia mater* from full text studies. In this approach, each extracted datapoint is associated with a particular text passage and the corresponding research paper. This allows the human reviewer to only focus on short pieces of text likely containing relevant data, and from the context, infer whether it is a primary measurement, a reference to a prior study, or referring to another context. If the extracted datapoint and the corresponding text passage suggest a primary mechanical study determined by the human reviewer, it is set aside for full manual review and possible inclusion in the SR. If instead the extracted data point originates from another paper that is likely to be a primary mechanical study which was not included in the initial article selection, it is also set aside for full manual review and possible inclusion in the SR.

Briefly, ChatExtract is a two-stage process, where sentences are first classified, and then expanded to short text passages used to extract structured data. A more detailed description of ChatExtract is given in section 2.3. Based on previous studies on mechanical properties, the classification step has close to 100% recall [59], while the subsequent data extraction and structuring step has a recall of over 87% [59,60]. Since the extraction and structuring step is validated manually after all primary studies are selected, ChatExtract effectively participates as a general classification step, ensuring an almost perfect recall in finding all relevant primary studies from the initial broad selection of papers. Compared to a fully manual selection process, full texts papers are presented to the reviewer as a condensed and focused set of data and short passages rather than as summarized abstracts which might not contain any relevant information. Furthermore, ChatExtract can screen a much larger volume of studies than would be viable for manual SR.

## 2.3) LLM-based Data Extraction Overview

A slightly modified version of ChatExtract was used. Originally, this method was designed to extract materials from research papers in the form of Material, Value, Unit triples (material triplets) for a given property; for the purpose of this study, the prompts were inverted to extract Property, Value, Unit triplets (property triplets) for a given material. The inverted process was more efficient than the original process because there were only 3 materials of interest, the *dura mater*, *arachnoid mater*, and *pia mater,* for which all material properties were to be reviewed. ChatExtract operates on short text passages composed of the paper title and two consecutive sentence chunks of the source text. To optimize the input to the LLM, a preprocessing stage used a set of regular expressions to exclude all passages that did not contain an isolated number (references, figure labels, table labels, and section numbers were not considered), meaning the excluded passage could not possibly contain any mechanical property data (Fig. 1). If appropriate, all relevant unit magnitudes (e.g., Pa, kPa, MPa, GPa) were included in the regular expression (Table S1). This limited the amount of text that needed to be processed by the LLM while ensuring no reduction of recall. The extracted data points included the value of interest, the corresponding passage, the article's title, and the digital object identifier (DOI). A fully detailed description can be found in the original ChatExtract publication [60]. The LLM used for the modified ChatExtract algorithm was GPT-4o with a temperature (randomness) of 0, frequency penalty of 0, presence penalty of 0, and all other parameters as default [60]. See section 2.1 in the supplemental information for validation on the choice of LLM. It is worth mentioning that the automatic portion of the review where LLMs are employed requires access to significant computational resources or a paid access to the LLM as a service.

## 2.4) Manual Data Extraction and Analysis

The extracted data points from the modified ChatExtract were structured into a table and manually checked by a human reviewer (BPC) for relevance. A property triplet was deemed relevant if the passage corresponded to the correct material and included data from a primary mechanical study. Adhering to the central principle of the snowball search method, which uses the reference list of a relevant article as a starting point to find additional related studies, property triplets from referenced values were manually screened (Fig. 1). For studies confirmed to contain relevant information, a human reviewer (BPC) read the entire paper to extract all mechanical properties and associated categorical and numerical data. Categorical data included species, sex, sample location, sample orientation, and study-specific treatment. Numerical data included sample dimensions, donor or animal age, post-mortem time, post-mortem temperature, test temperature, loading rate, sample preconditioning, mechanical results, and constitutive model parameters. Sample location was split into two parts: whether the sample was spinal or cranial, and the anatomical location relative to either the spine or cortex. For numerical data, the average, standard deviation, minimum, maximum, and median values were collected when available. When possible, data was manually extracted from relevant figures, with priority given to values stated in the text. The collected data was structured for visualization in a custom MATLAB plotting tool called BioAxis Studio, to visualize various trends across relevant studies. Structured data for the *dura mater, arachnoid mater,* and *pia mater* and source code for the BioAxis Studio are available for download on Github (https://github.com/HenakLab/Brain_Interface).

Extracted categorical and numerical data were analyzed in three different stages beginning with patterns emerging for a single variable across multiple studies. Studies that did not report a relevant value were excluded from this analysis. Interesting trends that arose from at least three studies were extended to investigate multi-variable affects using the BioAxis Studio tool. Finally, extracted studies and identified trends from the modified ChatExtract were compared to previous SRs.

**3) Results**

Data across all tissue types was summarized in comparison to other materials (Fig. 2). The *dura mater* and *pia mater* have similar mechanical properties to other biological materials and synthetic elastomers, while the *arachnoid mater* is most similar to synthetic hydrogels. Trends between tissues with respect to strain and elastic energy storage are visible. The *dura mater* has been the most extensively studied which is evident due to the higher density of data.

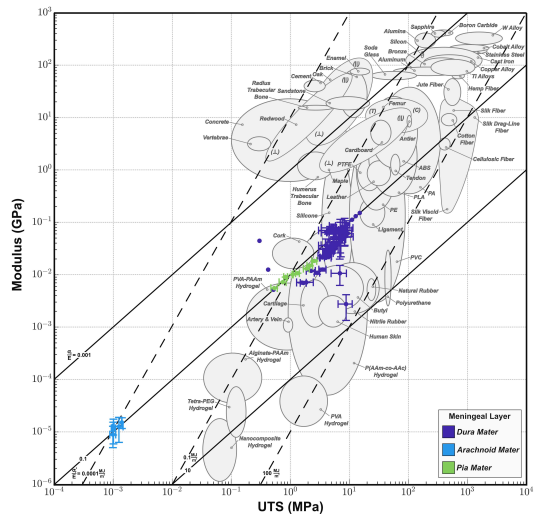

*Fig. 2: Ashby plot comparing the ultimate tensile strength (UTS) versus elastic modulus of the* dura mater (purple), arachnoid mater (blue), *and* pia mater (green) *in comparison to a wide range of other materials* [61,62]. *Guidelines at constant failure strain (solid black) and constant elastic energy storage (dashed black) are shown at relevant magnitudes for comparison between material groups.*

**3.1) Dura Mater**

*3.1.1) Search Space Results*

The initial search filter identified 17,951 studies corresponding to >1.8 million passages for the *dura mater* (Fig. 1). ChatExtract identified 3,707 passages as potentially relevant, which removed 99.79% of the total search space. Manual verification identified 58 relevant passages corresponding to 15 primary studies. Referenced values found an additional 32 studies, resulting in 47 of the original 17,951 studies ultimately included in the review. Compared to previous SR on the meninges, this review found 14 of the same studies identified by Pearcy et al. [49], who found 17 out of 400 studies related to elastic properties of the *dura mater*, and 15 of the same studies identified by Walsh et al. [50], who found 19 out of 663 studies related to *dura mater* mechanics. One of the studies included by Walsh et al. was identified by the modified ChatExtract algorithm, but was manually excluded by the reviewer due to lack of clarity in the reported mechanical values [50]. The 3 studies from Pearcy et al. [49] and the 4 studies from Walsh et al. [50] not identified in this review were not included in the initial search filter and therefore not considered by ChatExtract or the human reviewer. Accounting for duplicate studies, this review identified 28 studies related to *dura mater* biomechanics not previously included in an SR, 15 of which were published prior to Walsh et al. [50], and 16 of which were published prior to Pearcy et al. [49].

*3.1.2) Metadata: Study Characteristics and Reported Main Effects*

*Dura mater* studies varied widely in the reported details regarding sample information, testing configuration, environmental conditions and other metadata. Studies predominantly tested cranial tissue (32/47) and human tissue (27/47). Only a fraction of studies reported sample orientation (36/47), age (31/47), species (47/47), sex (22/47), post-mortem time (33/47), storage methods (41/47), test temperature/humidity (29/47), test hydration level (22/47), test type (45/47), and sample geometry (45/47). Full details on extracted data can be found in the *dura mater* master table in the supplemental information (Table S2).

Sample storage and post-mortem time varied considerably across all studies. Twenty-nine studies reported refrigerating or freezing samples with storage temperatures ranging from -80°C to 5°C, 22 reported storing samples in a physiological salt solution, and 14 studies utilized both methods. Post-mortem time, relevant to tissue degradation [63], was directly tested in 2 studies. These studies reported no significant change in mechanical properties up to 146 hours [26,64] for refrigerated [26], frozen [64], and hydrated [26,64] samples.

Tests were primarily performed in either uniaxial (31/47) or biaxial (6/47) tension. The effect of sample orientation was directly tested in 24 studies, finding preferred fiber orientations in spinal samples (14/24) and locally in cranial samples (6/24), but no preferred orientation globally for cranial samples (6/24).

Sex and age effects were less commonly tested or reported. Sex effects were tested in one study, which found no significant difference between maximum force, maximum tensile strength, or strain at maximum force [26]. Age effects indicated a decrease in elastic modulus throughout human lifespan [26,65], but no change between similar age cohorts [66–68].

Measurements of sample thickness were reported in 36 studies, and varied with respect to species, location, and age. The thickness of human cranial samples, ranging from 100 µm to 1880 µm (19/36), and human spinal samples, ranging from 100 µm to 275 µm (3/36), tended to increase with age [26,69] (Fig. 3A). The porcine animal model (Yorkshire (2/12), Landrace (2/12), and not reported (8/12)) was the most commonly used animal model. In porcine samples, cranial *dura mater* thickness, 214.90 µm to 1039 µm (6/36), and spinal *dura mater* thickness, 20 µm to 300 µm (8/36), tended to decrease with age [70–74] (Fig. 3B).

Constitutive models were fit to mechanical data in 22 studies including neo-Hookean (8/22), Ogden (6/22), and multi-fiber reinforced anisotropic models (4/22).

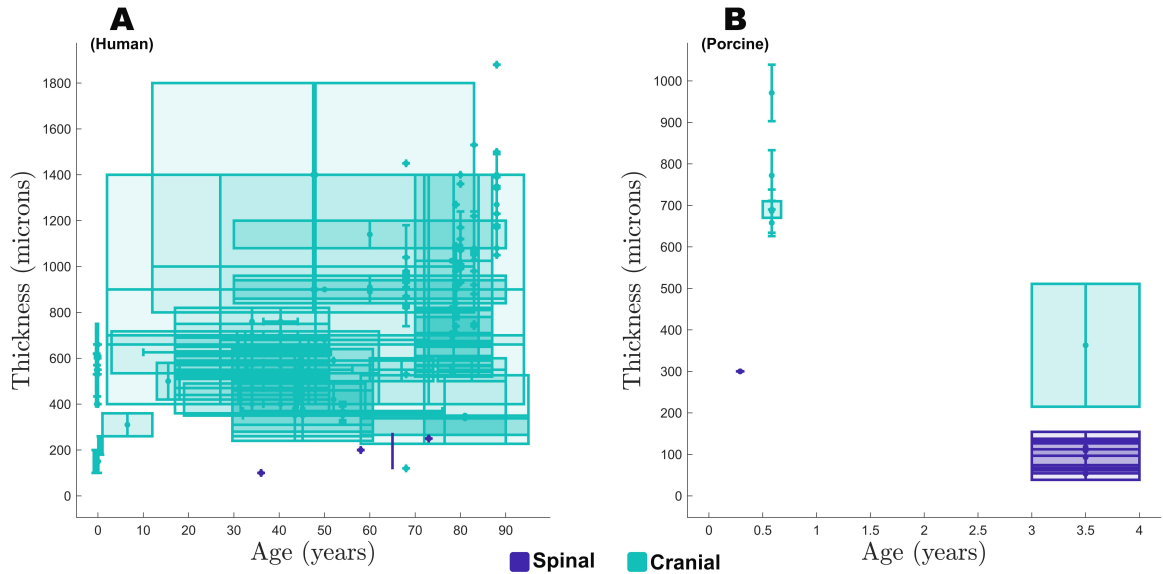

*Fig. 3. Effect of age on measured* dura mater *thickness in human (A) and porcine (B) samples show reverse trends, where thickness increases in human samples, but decreases in porcine samples with age.*

### *3.1.3)* Dura Mater *Mechanical Properties*

Qualitative trends between multiple variables were investigated using BioAxis Studio, and selected trends are highlighted here. Elastic modulus tended to increase with age for human cranial *dura mater*, ranging from 6.96 MPa for fetal tissue [67] to 62.6 MPa at 78.5 years [75] (Fig. 4A). Ultimate tensile stress and strain indicated an increasing ultimate tensile stress with age, from 1.57 MPa for fetal tissue [67] to 4.24 MPa at 78.5 years [75] (Fig. 4B), and a decreasing ultimate tensile strain, from 24% for fetal tissue [67] to 18% at 88 years [76] (Fig. 4C). No clear trend was visible for human spinal *dura mater* due to lack of data (Fig. 4A-C). However, porcine *dura mater* indicated three distinct sets of elastic moduli (Fig. 4D): spinal samples tested longitudinally (108.1 ± 40.3 MPa [71]), spinal samples tested transversely (53.2 ± 31.7 MPa [71]), and cranial samples (14.18 ± 2.48 MPa [77]).

For studies that fit an Ogden constitutive model, the Ogden stiffness ranged from -1.97 to 9.04 MPa and varied for location [78], orientation [79], and species [70] for some studies and not for others [64,70,71,80]. The Ogden exponentials for human tissue were generally negative (Fig.

S2), ranging from -94.33 to 65.56 and did not vary with location or orientation [80], while the Ogden exponentials for porcine tissue were generally positive (Fig. S2), ranging from 0.55 to 61.40 across multiple studies [64,70,71]. (Note that all Ogden parameters were transformed to a constant strain energy density prior to comparison, $W = \frac{\mu}{\alpha^2}(\lambda_1^\alpha + \lambda_2^\alpha + \lambda_3^\alpha)$, where $\mu$ is the stiffness, $\alpha$ is the exponential term, and $\lambda_i$ are the principal stretches.)

The *dura mater* exhibits complex rate dependence for elastic modulus and ultimate tensile stress, which can be grouped into three distinct regions (Fig. S3). The spread of Elastic modulus and ultimate tensile stress initially increase from quasi-static loading rates (10 mm/min) to dynamic loading rates (150 mm/min), followed by an overall decrease in magnitude at higher dynamic rates (1500 mm/min). At extreme dynamic loading rates (18,000 mm/min – 42,000 mm/min), the elastic modulus and ultimate tensile stress drop significantly. As the rate further increases, the ultimate tensile stress remains relatively constant while the elastic modulus begins to increase (Fig. S3).

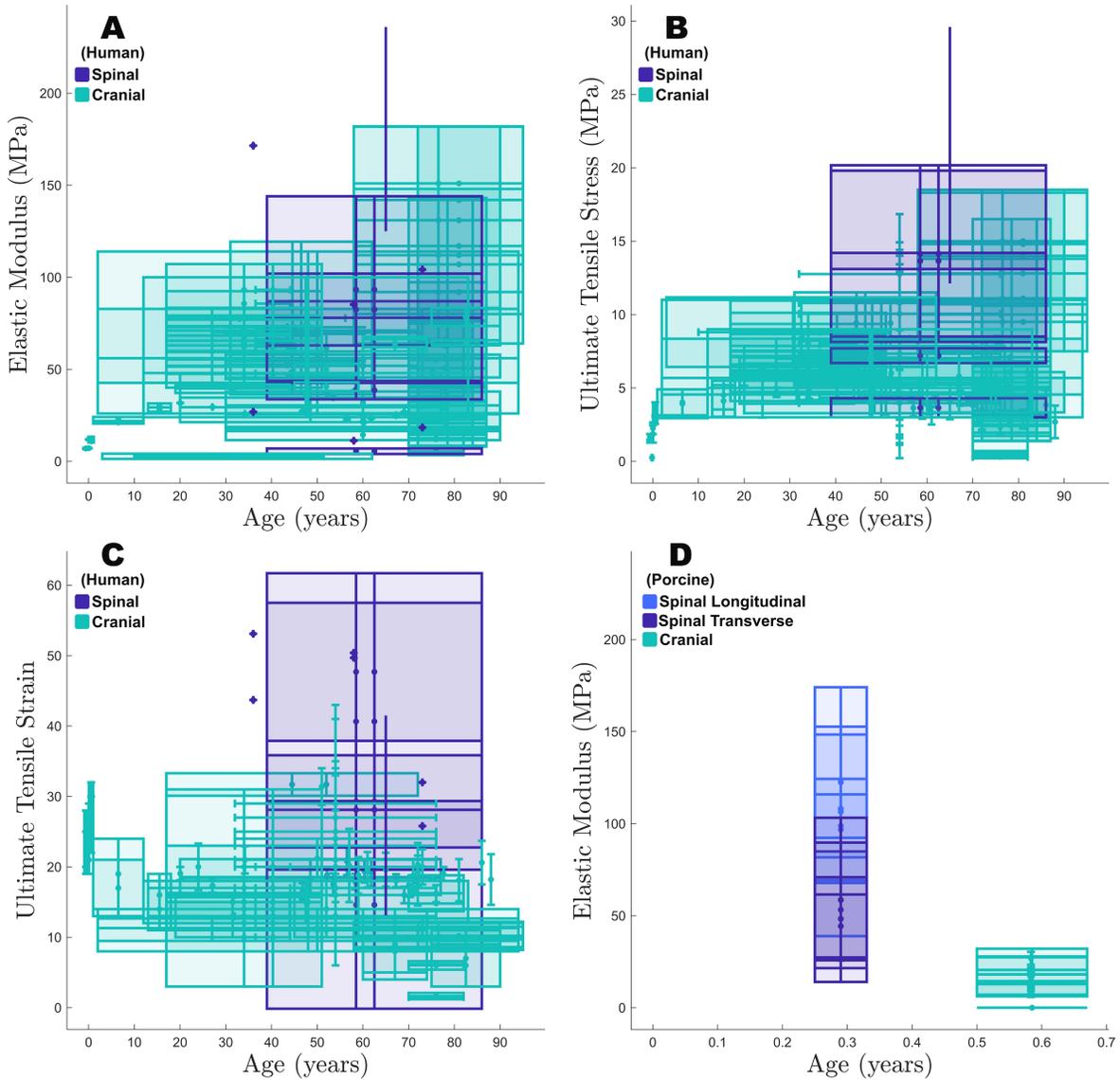

*Fig. 4.* Dura mater *mechanical properties for human samples show the effect of age on modulus (A), ultimate strength (B), and ultimate strain (C).* Porcine *dura mater* *elastic modulus results include orientation effects in spinal samples (D).*

### 3.2) Arachnoid Mater

### 3.2.1) Search Space Results

For the *arachnoid mater*, 1,816 studies were identified from the initial search filter across all 15 properties corresponding to over 230,000 passages (Fig. 1). ChatExtract identified 148 passages as potentially relevant, which removed 99.93% of the total search space. Following manual reclassification, 5 passages were deemed relevant corresponding to 4 primary mechanical studies. Referenced values led to an additional 4 studies, resulting in 8 of the 1,816 studies ultimately included in this review. Only Walsh et al. reviewed the pia-arachnoid complex (PAC) which referred to both the *arachnoid mater* and *pia mater* [50]. This review identified all 5 studies related to the PAC found by Walsh et al., 4 of which were manually classified as *arachnoid mater* based on sample information provided in the methods [50]. Accounting for duplicate studies, this review identified 3 new studies related to *arachnoid mater* biomechanics, where 1 study was published after Walsh et al. [50].

*3.1.2) Metadata: Study Characteristics and Reported Main Effects*

Despite the small number of studies, study characteristics varied, and were often not tested statistically for the *arachnoid mater*. Studies were primarily performed on cranial tissue (7/8) and bovine tissue (4/8). Human tissue was only tested in 1 study. Due to the low number of studies, specific information on sample orientation (1/8), age (8/8), sex (3/8), post-mortem time (8/8), species (8/8), storage methods (8/8), test temperature/humidity (2/8), test hydration level (7/8), test type (8/8), and sample geometry (7/8) was often reported without testing its effect. Full details can be found in the *arachnoid mater* master table in the supplemental information (Table S3).

In contrast to the *dura mater*, sample storage was consistent across all 8 studies. Samples were stored in either a physiological salt solution (3/8) or artificial CSF (5/8) and were mostly tested without a freeze-thaw cycle (7/8), within either a 6-hour (3/8) or 56-hour (8/8) period. Studies were reported as testing immature (4/8), mature (3/8), or elderly (1/8) tissue samples, with ages not comparable across studies of different species. Species of test samples included bovine (4/8), ovine (1/8), porcine (1/8), rat (1/8), and human (1/8). Tissue thickness ranged from 20 µm to 53 µm for human [81], bovine [82–85], and rat [86] tissue samples, while ovine [87] tissue samples ranged from 160 µm to 240 µm. Sex differences investigated for human tissue found no significant effect on tissue modulus, but found subject pair-wise differences for modulus, trabecular volume fraction, and membrane thickness [81]. Test methods included uniaxial tension (3/8), normal traction (2/8), uniaxial shear (2/8), atomic force microscopy (1/8), micro inflation (1/8), macro indentation (1/8), and micro indentation (1/8). Three studies used various constitutive models to fit mechanical data including a fiber reinforced viscoelastic Mooney-Rivlin [84], Maxwell and Ogden [86], and elastic thin-substrate constitutive models [88].

*3.1.3)* Arachnoid Mater *Mechanical Properties*

Any apparent effect of sample storage, post-mortem time, or sample age were overshadowed by differences between species (Fig. 5). Bovine *arachnoid mater* had the highest traction modulus [82,83,85], followed by porcine [88], human [81], and rat [86], which tended to scale with species body mass (Fig. 5). A power law fit to data averages resulted in an exponent of 0.54 with an $R^2$ of 0.957 and root mean square error normalized by the range of 7.54%. Related to tissue thickness, the volume fraction of trabeculae within the subarachnoid space varies considerably across the surface of the brain, leading to heterogenous mechanical properties for the *arachnoid mater* [81,86].

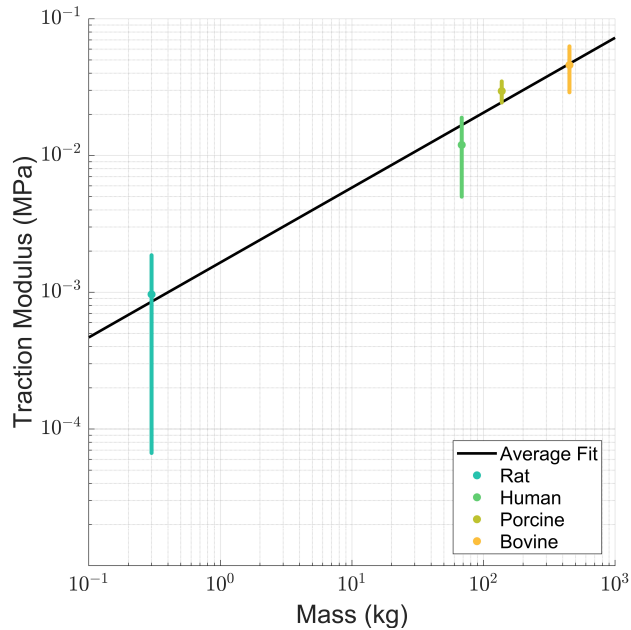

*Fig. 5:* Arachnoid mater *traction modulus across species as a function of body mass (Rat [89], Human [89], Porcine [90], Bovine [89]). The black line was directly fit to the four data averages which indicates a proportional increase in traction modulus with body size* $\texttt{Traction Modulus} = 0.00164 * \texttt{Mass}^{0.547}$ *with an $R^2$ of 0.957 and normalized root mean square error of 7.83%.*

### *3.3)* Pia Mater

### *3.2.1) Search Space Results*

For the *pia mater*, 5,654 studies were identified from the initial search filter across all 15 properties leading to over 600,000 passages (Fig. 1). ChatExtract deemed 375 passages as possibly relevant, removing 99.94% of the total search space. Through manual review, 12 passages corresponding to 3 primary mechanical papers were included. Tracing referenced values led to an additional 4 studies, resulting in 7 of the original 5,654 studies ultimately included in the review. Of the 5 studies found by Walsh et al. for the PAC, only 1 was classified as testing the *pia mater* for this review while the other 4 studies were classified as testing the *arachnoid mater [50]*. Accounting for duplicates, this review found 6 additional primary mechanical studies related to *pia mater* biomechanics, where 1 was published after Walsh et al. [50].

### *3.1.2) Metadata: Study Characteristics and Reported Main Effects*

*Pia mater* studies were tested exclusively in non-human samples. Other study characteristics were variable between studies. Studies were primarily performed on spinal tissue (5/7). While there was no human tissue, canine (1/7), bovine (1/7), ovine (1/7), lapine (1/7) and porcine *pia mater* (3/7) were tested. Specific information on sample orientation (4/7), age (4/7), sex (1/7), post-mortem time (7/7), species (7/7), storage methods (7/7), test temperature/humidity (5/7), test hydration level (7/7), test type (7/7), and sample geometry (7/7) were reported but not statistically evaluated. Full details can be found in the *pia mater* master table in the supplemental information (Table S4).

Similar to the *arachnoid mater*, sample storage for the *pia mater* was consistent across all relevant studies. Most studies stored samples in physiological salt solutions (7/8) between temperatures of 4°C to 24°C (5/7) and tested samples 16 hours post-mortem (7/7). Sample age varied with species, which were porcine (3/7), dog (1/7), lapine (1/7), bovine (1/7), and ovine (1/7), and were classified as immature (2/4) or mature (2/4). Significant effects of sample storage, post-mortem time, and sample age could not be investigated between species due to the low number of identified studies. Tissue thickness was reported in 3 studies, with ovine [28] ranging from 4.14 µm to 5.78 µm, rabbit [91] ranging from 9 µm to 15 µm, and porcine [92] ranging from 110 µm to 330 µm. Sex was only reported in one study [92], but its effects were not analyzed. Sample orientation was reported in 4 studies, 2 of which were on porcine spinal samples which found a higher elastic modulus and ultimate tensile strength in the longitudinal direction compared to the transverse direction [71,92]. The other 2 studies were only tested in one direction on different species [91,93]. Sample orientation was not investigated for cranial samples. Test methods included uniaxial tension (4/7), creep (2/7), biaxial (1/7), and stress relaxation (1/7). Various constitutive models were used to fit experimental data in five studies including a one term Ogden model [28,71,92] (3/5), neo-Hookean model [94] (1/5), and both isotropic and anisotropic Gasser-Ogden-Holzapfel (GOH) and Gasser models [71] (1/5).

*3.1.3) Pia Mater Mechanical Properties*

Because of the sparsity of reported data, trends were challenging to observe. The effect of sample thickness and species on mechanical properties could only be partially investigated based on reported data. Porcine [71,92,94] *pia mater* was the thickest and had the highest elastic modulus followed by lapine [91] and ovine [28] (Fig. 6A). Ultimate tensile stress tended to increase with ultimate tensile strain independent of species [28,71,92,94] (Fig. 6B). *Pia mater* behavior is nonlinear, with the elastic modulus increasing with strain rate [28,94] and strain [28,71,92,94].

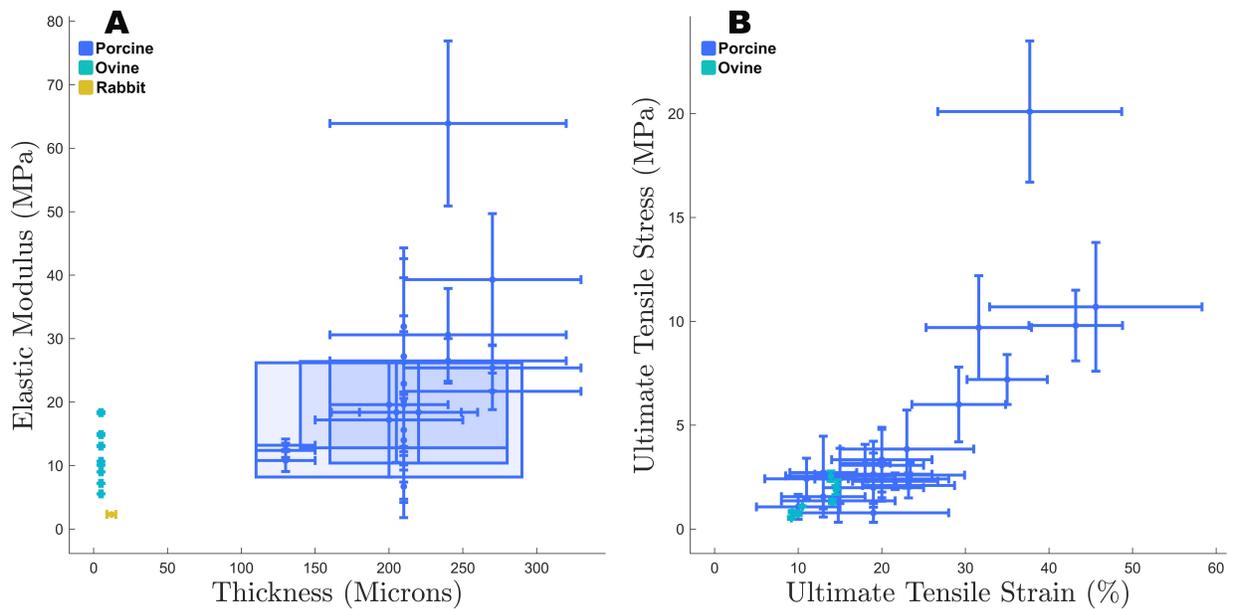

*Fig. 6:* Pia mater *thickness and elastic modulus across species (A), and ultimate tensile strain and ultimate tensile stress across species (B).*

## 4) Discussion

In this study, we conducted an LLM-assisted SR on the mechanical properties of the meninges. Each layer of the meninges was reviewed independently scanning 17,951 studies for the *dura mater*, 1,816 studies for the *arachnoid mater*, and 5,654 studies for the *pia mater*. Primary mechanical studies on meningeal mechanics were identified through a combination of an automated and manual review process resulting in 47 studies for the *dura mater*, 8 studies for the *arachnoid mater*, and 7 studies for the *pia mater*. Accounting for duplicates, this review identified 28 previously unreviewed studies related to *dura mater* mechanics, 3 studies related to *arachnoid mater* mechanics, and 6 studies related to *pia mater* mechanics.

Mechanical properties of biological tissues vary widely due to biological variation between species, anatomical locations, and individuals. The most notable source of variation for the meninges came from differences between species (e.g., bovine [82], human [27], ovine [28] , porcine [71], rabbit [91], and rat [86]). Variation in sample location, such as cranial tissue [28,73,81,95], spinal tissue [71,87,96], as well as anatomical location [43,70,71,77,81,82,87,97,98], contributed to variability as well. Tissue was most often tested in uniaxial tension, which commonly corresponded to reporting bulk elastic modulus, ultimate tensile strain, and ultimate tensile stress. The reported mechanical properties for the *dura* and *pia mater* are most similar to soft elastomers such as silicon, rubber, and other biological materials such as ligament and cartilage (Fig. 2). The *arachnoid mater* is noticeably softer and weaker than other biological materials and common synthetic hydrogels (Fig. 2). Material guidelines suggest similar elastic energy storage for the *dura mater* and *pia mater* with lower elastic energy storage for the *arachnoid mater,* but similar yield strains for each layer (Fig. 2). Compared to other materials, the meninges have comparable elastic energy storage to most engineered or hard materials with high yield strains most similar to synthetic elastomers (Fig. 2). While linear tissue properties are well reported [26,49,66,77], there remains a dearth of information on the nonlinear mechanical behavior, particularly for the *arachnoid* and *pia mater*. Descriptive mechanical studies investigating failure behavior and a wide range of loading rates are lacking; the effect of sample-specific factors such as age, sex, species, storage methods, and environmental test levels are similarly under reported. Reporting age in tandem with species is of critical importance for comparing animal models because different species age and mature at different rates. To best fill these gaps in the current literature, future studies should state species, sample location, sample orientation, sample treatment, sex, and any preconditioning and fully report mean, standard deviation, and range for age, post-mortem time, loading rate, sample length, sample width, and sample thickness.

The *dura mater* has been the most extensively studied meningeal tissue, and exhibits nonlinear [77], rate-dependent [42], viscoelastic [44] mechanical behavior, which is locally anisotropic and heterogenous in the skull [41,75], globally isotropic and homogenous in the skull [77,80], and globally anisotropic and heterogenous in the spine [71,92]. The most commonly reported constitutive models are a one term viscoelastic Ogden model for cranial tissue [64,80,95], and a viscoelastic GOH model, which considers longitudinal and transverse fiber orientations, for spinal tissue [71,75]. Specific model parameters vary significantly for species, age, and loading rate which require extra consideration when selecting constitutive model parameters for FE modeling. When possible, mechanical analysis should be conducted on human tissue hydrated in a physiological saline solution at body temperature [64,71,75,79,94]. In place of human tissue, the most commonly used animal model has been mature porcine tissue [43,64,70–74,77,96,98–100] as it most closely matches the *dura mater* thickness and elastic modulus (Fig. 3, 4A, 4D) as well as exhibiting similar lipid content [101,102], DNA polymerase [101–103], and water content [101] to human tissue during development [104]. Still unknown are the effects of tissue hydration post-mortem, dynamic loading rates (1000 – 100,000 mm/min), how relative age between species impacts tissue animal models, the transition point for isotropy and homogeneity for cranial tissue, and interactions between adjacent tissues.

Considerably less information exists on the mechanical properties for the *arachnoid* and *pia mater* compared to the *dura mater*. Bovine *arachnoid mater* has been extensively studied and modeled using a homogenous nonlinear, rate-dependent, viscoelastic constitutive model [82–85] but is considerably stiffer compared to other species including human tissue (Fig. 5). Additionally, this model did not consider structural nonlinearity arising from the multiple loading regimes of individual sub-arachnoid trabeculae [27], or the heterogeneity with respect to the skull [27,40]. Of all animal models reported, porcine tissue most closely resembled the elastic modulus [88] and thickness [88] of human *arachnoid mater* (Fig. 5), though it has a higher volume fraction of sub arachnoid trabeculae [39,40]. Anatomical studies [30,32,105] suggest that testing the *arachnoid* in isolation from the *pia* is an arduous task due to the anchoring of the subarachnoid trabeculae in the *pia mater*. Future studies on the mechanical properties of the *arachnoid mater* should be conducted on hydrated human or porcine tissue with the *pia mater* intact. Studies on human *pia mater* were not found in this review, and spinal and cranial *pia mater* has not been tested with in the same species. Constitutive models fit to *pia mater* experimental data found loading rate dependent [28,94], nonlinear [106], viscoelastic [93] behavior that varies in direction and location along the spine [71,92]. The *pia mater* can be tested in isolation from the *arachnoid mater*; however, a process for isolating the *pia mater* from the *arachnoid mater* without damage is still unknown. Current FE head models often use a simplified linear elastic constitutive model for the *pia mater* with a modulus of 2.3 MPa [91]; however, multiple studies in this review have reported elastic moduli an order of magnitude higher in both the longitudinal (~ 20 MPa) and transverse (~10 MPa) directions [28,71,92] (Fig.

6A). Future studies should investigate how species, age, sex, loading rate, post-mortem time, and tissue hydration affect the complex mechanical behavior of the *pia mater*.

The semi-automatic approach to a literature review required less time to process a larger number of studies. While exact hours were not tracked, the manual portion of the review took between 600 - 680 hours to complete (corresponding to 15 - 17 standard 40-hr work weeks) to review over 25,000 studies. A standard SR which averages 4.34 studies per person hour (881 person-hours to review 3781 studies [54]) would require over 100,000 hours to review the same 25,000 studies. While this review included a much larger breadth of studies compared to a standard SR, there are some limitations in this analysis. Only full text studies from ScienceDirect that were open source or available under the UW-Madison license were considered. The majority of studies (>99.9%) were automatically scanned using a conversational LLM where it is difficult to know what factors were considered when reviewing studies. Manual evaluation was only applied after automatic classification, which could have led to the rejection of some relevant studies without the reviewer's knowledge. This effect was minimized by using a combination of general and specific search terms which was shown to have different levels of precision and recall (Fig. S1). The automatic portion of the review required access to considerable computational resources or a paid access to the LLM as a service. Additionally, only a single reviewer (BPC) analyzed the results, which could lead to bias in data extraction, classification, and relevancy.

Current FE head models often neglect or oversimplify the mechanical properties of the meninges, which can drastically influence prediction accuracy. This study conducted a semi-automatic systematic review of the mechanical properties of the meninges using a modified ChatExtract algorithm and a single human reviewer. Over 25,000 papers (~2.6 million passages) were considered where 47 studies related to the *dura mater*, 8 studies related to the *arachnoid mater*, and 7 studies related to the *pia mater* were ultimately included. All three layers of the meninges were found to be rate dependent, nonlinear viscoelastic and individually vary with respect to species, age, location, and orientation. Future FE head models that use a linear elastic constitutive model for the *pia mater* should use an order of magnitude higher elastic modulus compared to previous FE models. Studies on the interaction between different meningeal layers were sparse which could lead to inaccurate modeling of tissue boundaries even with accurate constitutive models. Future research on meningeal mechanical properties should primarily investigate a broader spectrum of quasi-static and dynamic loading rates, considering both the presence and absence of age-related effects on the *arachnoid* and *pia mater*, as these factors are anticipated to influence the accuracy of finite element (FE) predictions.

## 5) Acknowledgements

We gratefully acknowledge support from the Office of Naval Research via the PANTHER program under Dr. Timothy Bentley (N00014-23-1-2825), and the National Science Foundation (NSF) Cyberinfrastructure for Sustained Scientific Innovation (CSSI) (1931298). This project would not have been possible without the technical and administrative support provided by librarian Dave Bloom in database selection and review methodology as well as Nolan Thole for LLM and search term validation. Their efforts were crucial to handling such a large number of studies and for a smooth execution of this project.

## 6) Declaration of Competing Interest

The authors declare that they have no known competing financial interests or personal relationships that could have appeared to influence the work reported in this paper.

## 7) CRediT authorship contribution statement

**Brandon P. Chelstrom:** Methodology, Software, Validation, Investigation, Data Curation, Writing – Original Draft, Review, & Editing, Visualization **Maciej P. Polak:** Methodology, Software, Validation, Writing – Original Draft, Review, & Editing **Dane Morgan:** Conceptualization, Funding Acquisition, Writing – Review, & Editing **Corinne R. Henak:** Conceptualization, Resources, Writing – Review, & Editing, Supervision, Funding Acquisition

## 8) References

[1] J. Daugherty, D. Waltzman, K. Sarmiento, L. Xu, Morbidity and Mortality Weekly Report Traumatic Brain Injury-Related Deaths by Race/Ethnicity, Sex, Intent, and Mechanism of Injury-United States, 2000-2017, 2019.
https://www.cdc.gov/nchs/products/databriefs/db328.htm.

[2] J. Daugherty, K. Sarmiento, D. Waltzman, L. Xu, Traumatic Brain Injury–Related Hospitalizations and Deaths in Urban and Rural Counties—2017, Ann Emerg Med 79 (2022) 288-296.e1. https://doi.org/10.1016/j.annemergmed.2021.09.433.

[3] J.A. Langlois, W. Rutland-Brown, M.M. Wald, The Epidemiology and Impact of Traumatic Brain Injury A Brief Overview, J Head Trauma Rehabil 21 (2006) 375–378.
http://journals.lww.com/headtraumarehab.

[4] D. Thurman, C. Alverson, D. Browne, K. Dunn, J. Guerrero, R. Johnson, V. Johnson, J. Langlois, D. Pilkey, J. Sniezek, S. Toal, P. Gosler, B. Gabella, R. Hoffman, G. Whieneck, M. Kinde, J. Roesler, G. Land, M. Tuinen, L. Santilli, K. Thoburn, G. Feck, P. Archer, M. Crutcher, S. Mallonee, E. McCutcheon, A. Selassie, L. Frazier, Traumatic Brain Injury in the United States: A Report to Congress, 1999.

[5] The Management and Rehabilitation of Post-Acute Mild Traumatic Brain Injury Work Group, VA/DoD Clinical Practice Guideline for the Management and Rehabilitation of Post-Acute Mild Traumatic Brain Injury, 2021. www.tricare.mil.

[6] S.M. Slobounov, A. Walter, H.C. Breiter, D.C. Zhu, X. Bai, T. Bream, P. Seidenberg, X. Mao, B. Johnson, T.M. Talavage, The effect of repetitive subconcussive collisions on brain integrity in collegiate football players over a single football season A multi-modal neuroimaging study, Neuroimage Clin 14 (2017) 708–718.
https://doi.org/10.1016/j.nicl.2017.03.006.

[7] T.M. Talavage, E.A. Nauman, E.L. Breedlove, U. Yoruk, A.E. Dye, K.E. Morigaki, H. Feuer, L.J. Leverenz, Functionally-detected cognitive impairment in high school football players without clinically-diagnosed concussion, J Neurotrauma 31 (2014) 327–338.
https://doi.org/10.1089/neu.2010.1512.

[8] S. Parikh, M. Koch, R.K. Narayan, Traumatic Brain Injury, Int Anesthesiol Clin 45 (2007) 119–135. http://journals.lww.com/anesthesiaclinics.

[9] A.G. Kolias, P.J. Kirkpatrick, P.J. Hutchinson, Decompressive craniectomy: past, present and future, Nat Rev Neurol 9 (2013) 405–415. https://doi.org/10.1038/nrneurol.2013.106.

[10] J.S. Giudice, W. Zeng, T. Wu, A. Alshareef, D.F. Shedd, M.B. Panzer, An Analytical Review of the Numerical Methods used for Finite Element Modeling of Traumatic Brain Injury, Ann Biomed Eng 47 (2019) 1855–1872. https://doi.org/10.1007/s10439-018-02161-5.

[11] N.G. Ibrahim, R. Natesh, S.E. Szczesny, K. Ryall, S.A. Eucker, B. Coats, S.S. Margulies, In situ deformations in the immature brain during rapid rotations, J Biomech Eng 132 (2010). https://doi.org/10.1115/1.4000956.

[12] R.W. Carlsen, A.L. Fawzi, Y. Wan, H. Kesari, C. Franck, A quantitative relationship between rotational head kinematics and brain tissue strain from a 2-D parametric finite element analysis, Brain Multiphys 2 (2021). https://doi.org/10.1016/j.brain.2021.100024.

[13] N.C. Colgan, M.D. Gilchrist, K.M. Curran, Applying DTI white matter orientations to finite element head models to examine diffuse TBI under high rotational accelerations,

Prog Biophys Mol Biol 103 (2010) 304–309. https://doi.org/10.1016/j.pbiomolbio.2010.09.008.

[14] B. Coats, G. Binenbaum, C. Smith, R.L. Peiffer, C.W. Christian, A.C. Duhaime, S.S. Margulies, Cyclic head rotations produce modest brain injury in infant piglets, J Neurotrauma 34 (2017) 235–247. https://doi.org/10.1089/neu.2015.4352.

[15] A. Al-Bsharat, W. Hardy, K. Yang, T. Khalil, Brain/Skull Relative Displacement Magnitude Due to Blunt Head Impact: New Experimental Data and Model, Stamp Car Crash Conference Proceedings (1999) 321–331.

[16] K.A. Kailash, C.A. Guertler, C.L. Johnson, R.J. Okamoto, P. V. Bayly, Measurement of relative motion of the brain and skull in the mini-pig in-vivo, J Biomech 156 (2023). https://doi.org/10.1016/j.jbiomech.2023.111676.

[17] A.A. Badachhape, R.J. Okamoto, R.S. Durham, B.D. Efron, S.J. Nadell, C.L. Johnson, P. V. Bayly, The Relationship of Three-Dimensional Human Skull Motion to Brain Tissue Deformation in Magnetic Resonance Elastography Studies, J Biomech Eng 139 (2017). https://doi.org/10.1115/1.4036146.

[18] M.M. Mortazavi, S.A. Quadri, M.A. Khan, A. Gustin, S.S. Suriya, T. Hassanzadeh, K.M. Fahimdanesh, F.H. Adl, S.A. Fard, M.A. Taqi, I. Armstrong, B.A. Martin, R.S. Tubbs, Subarachnoid Trabeculae: A Comprehensive Review of Their Embryology, Histology, Morphology, and Surgical Significance, World Neurosurg 111 (2018) 279–290. https://doi.org/10.1016/j.wneu.2017.12.041.

[19] G.G. Scott, S.S. Margulies, B. Coats, Utilizing multiple scale models to improve predictions of extra-axial hemorrhage in the immature piglet, Biomech Model Mechanobiol 15 (2016) 1101–1119. https://doi.org/10.1007/s10237-015-0747-0.

[20] S. Yang, J. Tang, B. Nie, Q. Zhou, Assessment of brain injury characterization and influence of modeling approaches, Sci Rep 12 (2022). https://doi.org/10.1038/s41598-022-16713-2.

[21] S. Sullivan, S.A. Eucker, D. Gabrieli, C. Bradfield, B. Coats, M.R. Maltese, J. Lee, C. Smith, S.S. Margulies, White matter tract-oriented deformation predicts traumatic axonal brain injury and reveals rotational direction-specific vulnerabilities, Biomech Model Mechanobiol 14 (2015) 877–896. https://doi.org/10.1007/s10237-014-0643-z.

[22] L. Zhang, K.H. Yang, A.I. King, A Proposed Injury Threshold for Mild Traumatic Brain Injury, J Biomech Eng 126 (2004) 226–236. https://doi.org/10.1115/1.1691446.

[23] E. Bar-Kochba, M.T. Scimone, J.B. Estrada, C. Franck, Strain and rate-dependent neuronal injury in a 3D in vitro compression model of traumatic brain injury, Sci Rep 6 (2016). https://doi.org/10.1038/srep30550.

[24] H. Gray, P. Williams, L. Bannister, Gray's anatomy, 38th ed., Churchill Livingstone, Edinburgh, 1995.

[25] F. Herisson, V. Frodermann, G. Courties, D. Rohde, Y. Sun, K. Vandoorne, G.R. Wojtkiewicz, G.S. Masson, C. Vinegoni, J. Kim, D.E. Kim, R. Weissleder, F.K. Swirski, M.A. Moskowitz, M. Nahrendorf, Direct vascular channels connect skull bone marrow and the brain surface enabling myeloid cell migration, Nat Neurosci 21 (2018) 1209–1217. https://doi.org/10.1038/s41593-018-0213-2.

[26] J. Zwirner, M. Scholze, J.N. Waddell, B. Ondruschka, N. Hammer, Mechanical Properties of Human Dura Mater in Tension – An Analysis at an Age Range of 2 to 94 Years, Sci Rep 9 (2019). https://doi.org/10.1038/s41598-019-52836-9.

[27]   N. Benko, E. Luke, Y. Alsanea, B. Coats, Mechanical characterization of the human pia-arachnoid complex, J Mech Behav Biomed Mater 120 (2021). https://doi.org/10.1016/j.jmbbm.2021.104579.

[28]   Y. Li, W. Zhang, Y.C. Lu, C.W. Wu, Hyper-viscoelastic mechanical behavior of cranial pia mater in tension, Clinical Biomechanics 80 (2020). https://doi.org/10.1016/j.clinbiomech.2020.105108.

[29]   F. Vandenabeele, J. Creemers, I. Lambrichts, Ultrastructure of the human spinal arachnoid mater and dura mater, J. Anat 189 (1996) 417–430.

[30]   R.O. Weller, Microscopic morphology and histology of the human meninges, Morphologie 89 (2005) 22–34. https://doi.org/10.1016/S1286-0115(05)83235-7.

[31]   R. Alcolado, R.O. Weller, P. Parrish, D. Garrod, The Cranial Arachnoid and Pia Mater in Man: Anatomical and Ultrastructural Observations, Neuropathol Appl Neurobiol 14 (1988) 1–17. https://doi.org/10.1111/j.1365-2990.1988.tb00862.x.

[32]   P. Saboori, A. Sadegh, Histology and Morphology of the Brain Subarachnoid Trabeculae, Anat Res Int 2015 (2015) 1–9. https://doi.org/10.1155/2015/279814.

[33]   C.W. Kerber, T.H. Newton, The Macro and Microvasculature of the Dura Mater*, Neuroradiology 6 (1973) 175–179.

[34]   M. Protasoni, S. Sangiorgi, A. Cividini, G.T. Culuvaris, G. Tomei, C. Dell'Orbo, M. Raspanti, S. Balbi, M. Reguzzoni, The collagenic architecture of human dura mater: Laboratory investigation, J Neurosurg 114 (2011) 1723–1730. https://doi.org/10.3171/2010.12.JNS101732.

[35]   W. Schachenmayr, R.L. Friede, The Origin of Subdural Neomembranes I. Fine Structure of the Dura-Arachnoid Interface in Man, American Journal of Pathology (1978) 53–62.

[36]   N. Adeeb, M.M. Mortazavi, R.S. Tubbs, A.A. Cohen-Gadol, The cranial dura mater: A review of its history, embryology, and anatomy, Child's Nervous System 28 (2012) 827–837. https://doi.org/10.1007/s00381-012-1744-6.

[37]   N. Adeeb, M.M. Mortazavi, A. Deep, C.J. Griessenauer, K. Watanabe, M.M. Shoja, M. Loukas, R.S. Tubbs, The pia mater: A comprehensive review of literature, Child's Nervous System 29 (2013) 1803–1810. https://doi.org/10.1007/s00381-013-2044-5.

[38]   A. Kinaci, W. Bergmann, R.L.A.W. Bleys, A. van der Zwan, T.P.C. van Doormaal, Histologic comparison of the dura mater among species, Comp Med 70 (2020) 170–175. https://doi.org/10.30802/AALAS-CM-19-000022.

[39]   G.G. Scott, B. Coats, Microstructural Characterization of the Pia-Arachnoid Complex Using Optical Coherence Tomography, IEEE Trans Med Imaging 34 (2015) 1452–1459. https://doi.org/10.1109/TMI.2015.2396527.

[40]   N. Benko, E. Luke, Y. Alsanea, B. Coats, Spatial distribution of human arachnoid trabeculae, J Anat 237 (2020) 275–284. https://doi.org/10.1111/joa.13186.

[41]   M.S. Sacks, J. Hamann, S.E. Otano-Lata, T.I. Malinin, Local Mechanical Anisotropy in Human Cranial Dura Mater Allografts, J Biomech Eng 120 (1998) 541–544. http://asmedigitalcollection.asme.org/biomechanical/article-pdf/120/4/541/5767037/541_1.pdf.

[42]   J. Zwirner, B. Ondruschka, M. Scholze, A. Thambyah, J. Workman, N. Hammer, J.A. Niestrawska, Dynamic load response of human dura mater at different velocities, J Mech Behav Biomed Mater 138 (2023). https://doi.org/10.1016/j.jmbbm.2022.105617.

[43] E. Mazgajczyk, K. Scigala, M. Czy, W. Jarmundowicz, R. BedZinski, Mechanical Properties of Cervical Dura Mater, Acta of Bioengineering and Biomaterials 14 (2012) 51–58.

[44] J. Galford E., J. McElhaney H., A Viscoelastic Study of Scalp, Brain, and Dura, J Biomech 3 (1970) 211–221. https://doi.org/10.1016/0021-9290(70)90007-2.

[45] R.K. Wilcox, L.E. Bilston, D.C. Barton, R.M. Hall, Mathematical model for the viscoelastic properties of dura mater, in: J Orthop Sci, 2003: pp. 432–434.

[46] L. Gu, M.S. Chafi, S. Ganpule, N. Chandra, The influence of heterogeneous meninges on the brain mechanics under primary blast loading, Compos B Eng 43 (2012) 3160–3166. https://doi.org/10.1016/j.compositesb.2012.04.014.

[47] A. Madhukar, M. Ostoja-Starzewski, Finite Element Methods in Human Head Impact Simulations: A Review, Ann Biomed Eng 47 (2019) 1832–1854. https://doi.org/10.1007/s10439-019-02205-4.

[48] K. Ming Tse, H. Lee, S. Piang Lim, V. Beng Chye Tan, H. Pueh Lee, H.A. Pueh Lee, A Review of Head Injury and Finite Element Head Models, American Journal of Engineering, Technology, and Society 1 (2014) 28–52. http://www.openscienceonline.com/journal/ajets.

[49] Q. Pearcy, J. Tomlinson, J.A. Niestrawska, D. Möbius, M. Zhang, J. Zwirner, Systematic review and meta-analysis of the biomechanical properties of the human dura mater applicable in computational human head models, Biomech Model Mechanobiol 21 (2022) 755–770. https://doi.org/10.1007/s10237-022-01566-5.

[50] D.R. Walsh, Z. Zhou, X. Li, J. Kearns, D.T. Newport, J.J.E. Mulvihill, Mechanical Properties of the Cranial Meninges: A Systematic Review, J Neurotrauma 38 (2021) 1748–1761. https://doi.org/10.1089/neu.2020.7288.

[51] M.J. Page, J.E. McKenzie, P.M. Bossuyt, I. Boutron, T.C. Hoffmann, C.D. Mulrow, L. Shamseer, J.M. Tetzlaff, E.A. Akl, S.E. Brennan, R. Chou, J. Glanville, J.M. Grimshaw, A. Hróbjartsson, M.M. Lalu, T. Li, E.W. Loder, E. Mayo-Wilson, S. McDonald, L.A. McGuinness, L.A. Stewart, J. Thomas, A.C. Tricco, V.A. Welch, P. Whiting, D. Moher, The PRISMA 2020 statement: An updated guideline for reporting systematic reviews, The BMJ 372 (2021). https://doi.org/10.1136/bmj.n71.

[52] R.A. Damarell, N. May, S. Hammond, R.M. Sladek, J.J. Tieman, Topic search filters: a systematic scoping review, Health Info Libr J 36 (2019) 4–40. https://doi.org/10.1111/hir.12244.

[53] V.J. White, J.M. Glanville, C. Lefebvre, T.A. Sheldon, A statistical approach to designing search filters to find systematic reviews: objectivity enhances accuracy, J Inf Sci (2001). https://doi.org/https://doi.org/10.1177/016555150102700601.

[54] B. Pham, E. Bagheri, P. Rios, A. Pourmasoumi, R.C. Robson, J. Hwee, W. Isaranuwatchai, N. Darvesh, M.J. Page, A.C. Tricco, Improving the conduct of systematic reviews: a process mining perspective, J Clin Epidemiol 103 (2018) 101–111. https://doi.org/10.1016/j.jclinepi.2018.06.011.

[55] L. Bornmann, R. Haunschild, R. Mutz, Growth rates of modern science: a latent piecewise growth curve approach to model publication numbers from established and new literature databases, Humanit Soc Sci Commun 8 (2021). https://doi.org/10.1057/s41599-021-00903-w.

[56]    S.R. Jonnalagadda, P. Goyal, M.D. Huffman, Automating data extraction in systematic reviews: A systematic review, Syst Rev 4 (2015). https://doi.org/10.1186/s13643-015-0066-7.

[57]    H. Khalil, D. Ameen, A. Zarnegar, Tools to support the automation of systematic reviews: a scoping review, J Clin Epidemiol 144 (2022) 22–42. https://doi.org/10.1016/j.jclinepi.2021.12.005.

[58]    Q. Khraisha, S. Put, J. Kappenberg, A. Warraitch, K. Hadfield, Can large language models replace humans in systematic reviews? Evaluating GPT-4's efficacy in screening and extracting data from peer-reviewed and grey literature in multiple languages, Res Synth Methods (2024). https://doi.org/10.1002/jrsm.1715.

[59]    M.P. Polak, S. Modi, A. Latosinska, J. Zhang, C.W. Wang, S. Wang, A.D. Hazra, D. Morgan, Flexible, model-agnostic method for materials data extraction from text using general purpose language models, Digital Discovery 3 (2024) 1221–1235. https://doi.org/10.1039/d4dd00016a.

[60]    M.P. Polak, D. Morgan, Extracting Accurate Materials Data from Research Papers with Conversational Language Models and Prompt Engineering, (2023). http://arxiv.org/abs/2303.05352.

[61]    A. Li, Y. Si, X. Wang, X. Jia, X. Guo, Y. Xu, Poly(vinyl alcohol) Nanocrystal-Assisted Hydrogels with High Toughness and Elastic Modulus for Three-Dimensional Printing, ACS Appl Nano Mater 2 (2019) 707–715. https://doi.org/10.1021/acsanm.8b01786.

[62]    M. Ashby F., Materials Selection in Mechanical Design, 5th ed., Todd Green, 2016.

[63]    J. Exton, J.M.G. Higgins, J. Chen, Acute brain slice elastic modulus decreases over time, Sci Rep 13 (2023) 12826. https://doi.org/10.1038/s41598-023-40074-z.

[64]    B. Pierrat, L. Carroll, F. Merle, D.B. MacManus, R. Gaul, C. Lally, M.D. Gilchrist, A. Ní Annaidh, Mechanical Characterization and Modeling of the Porcine Cerebral Meninges, Front Bioeng Biotechnol 8 (2020). https://doi.org/10.3389/fbioe.2020.00801.

[65]    R. Van Noort, T.R.P. Martin, M.M. Black, A.T. Barker, C.G. Montero, The mechanical properties of human dura mater and the effects of storage media, Clin. Phys. Physiol. Meas 2 (1981) 197–203. https://doi.org/https://doi.org/10.1088/0143-0815/2/3/003.

[66]    R. Van Noort, M.M. Black, T.R.P. Martin, S. Meanley, A study of the uniaxial mechanical properties of human dura mater preserved in glycerol, 1980.

[67]    V.I. Zyablov, Y.N. Shapovalov, K.D. Toskin, V. V Tkach, V. V Zhebrovsky, L.S. Georgievskaya, Structure and Physical-Mechanical Properties of the Human Dural Mind in the Age Aspect, Histology and Embryology 132 (1982) 29–35.

[68]    Q. Pearcy, M. Jeejo, M. Scholze, J. Tomlinson, J. Dressler, M. Zhang, J. Zwirner, Biomechanics of vascular cranial areas of the human cranial dura mater, J Mech Behav Biomed Mater 125 (2022). https://doi.org/10.1016/j.jmbbm.2021.104866.

[69]    E. Zarzur, Mechanical Properties of the Human Lumbar Dura Mater, Arq Neuropsiquiatr 54 (1996) 455–460.

[70]    S. Cavelier, R.D. Quarrington, C.F. Jones, Mechanical properties of porcine spinal dura mater and pericranium, J Mech Behav Biomed Mater 126 (2022). https://doi.org/10.1016/j.jmbbm.2021.105056.

[71]    M. Evin, P. Sudres, P. Weber, Y. Godio-Raboutet, P.-J. Arnoux, E. Wagnac, Y. Petit, Y. Tillier, Experimental Bi-axial tensile tests of spinal meningeal tissues and constitutive models comparison, Acta Biomater 140 (2022) 446–456.

[72] A. Sharma, E. Moore, L.N. Williams, An in vitro study of micromechanics, cellular proliferation and viability on both decellularized porcine dura grafts and native porcine dura grafts, Biomedical Engineering Advances 6 (2023) 100108. https://doi.org/10.1016/j.bea.2023.100108.

[73] Y. Su, Z. Li, H. Zhu, J. He, B. Wei, D. Li, Electrohydrodynamic Fabrication of Triple-layered Polycaprolactone Dura Mater Substitute with Antibacterial and Enhanced Osteogenic Capability, Chinese Journal of Mechanical Engineering: Additive Manufacturing Frontiers 1 (2022) 100026. https://doi.org/10.1016/j.cjmeam.2022.100026.

[74] A. Sharma, J. Liao, L.N. Williams, Structure and mechanics of native and decellularized porcine cranial dura mater, Engineered Regeneration 4 (2023) 205–213. https://doi.org/10.1016/j.engreg.2023.02.004.

[75] J.A. Niestrawska, M. Rodewald, C. Schultz, E. Quansah, T. Meyer-Zedler, M. Schmitt, J. Popp, I. Tomasec, B. Ondruschka, N. Hammer, Morpho-mechanical mapping of human dura mater microstructure, Acta Biomater 170 (2023) 86–96. https://doi.org/10.1016/j.actbio.2023.08.024.

[76] V.N. R, M. T.R.P., B. M.M., B. A.T., C.G. Montero, The Mechanical Propeties of Human Dura Mater and the Effects of Storage Media, Clin. Phys. Physiol. Meas 2 (1981) 197–203.

[77] D.R. Walsh, A.M. Ross, S. Malijauskaite, B.D. Flanagan, D.T. Newport, K.D. McGourty, J.J.E. Mulvihill, Regional mechanical and biochemical properties of the porcine cortical meninges, Acta Biomater 80 (2018) 237–246. https://doi.org/10.1016/j.actbio.2018.09.004.

[78] J.T. Maikos, R.A.I. Elias, D.I. Shreiber, Mechanical properties of dura mater from the rat brain and spinal cord, J Neurotrauma 25 (2008) 38–51. https://doi.org/10.1089/neu.2007.0348.

[79] C. Persson, S. Evans, R. Marsh, J.L. Summers, R.M. Hall, Poisson's ratio and strain rate dependency of the constitutive behavior of spinal dura mater, Ann Biomed Eng 38 (2010) 975–983. https://doi.org/10.1007/s10439-010-9924-6.

[80] D. De Kegel, J. Vastmans, H. Fehervary, B. Depreitere, J. Vander Sloten, N. Famaey, Biomechanical characterization of human dura mater, J Mech Behav Biomed Mater 79 (2018) 122–134. https://doi.org/10.1016/j.jmbbm.2017.12.023.

[81] N. Benko, E. Luke, Y. Alsanea, B. Coats, Mechanical characterization of the human pia-arachnoid complex, J Mech Behav Biomed Mater 120 (2021). https://doi.org/10.1016/j.jmbbm.2021.104579.

[82] X. Jin, C. Ma, L. Zhang, K.H. Yang, A.I. King, G. Dong, J. Zhang, Biomechanical Response of the Bovine Pia-Arachnoid Complex to Normal Traction Loading at Varying Strain Rates, 2007.

[83] X. Jin, K.H. Yang, A.I. King, Mechanical properties of bovine pia-arachnoid complex in shear, J Biomech 44 (2011) 467–474. https://doi.org/10.1016/j.jbiomech.2010.09.035.

[84] X. Jin, H. Mao, K.H. Yang, A.I. King, Constitutive modeling of pia-arachnoid complex, Ann Biomed Eng 42 (2014) 812–821. https://doi.org/10.1007/s10439-013-0948-6.

[85] X. Jin, J. Lee B., L. Leung Yee, L. Zhang, K. Yang H., A. King I., Biomechanical response of the Bovine Pia-Arachnoid Complex to Tensile Loading at Varying Strain-Rates, Stapp Car Crash J 50 (2006) 637–649. https://doi.org/https://doi.org/10.4271/2006-22-0025.

[86]    G. Fabris, Z. M. Suar, M. Kurt, Micromechanical heterogeneity of the rat pia-arachnoid complex, Acta Biomater 100 (2019) 29–37. https://doi.org/10.1016/j.actbio.2019.09.044.

[87]    N.L. Ramo, K.L. Troyer, C.M. Puttlitz, Viscoelasticity of spinal cord and meningeal tissues, Acta Biomater 75 (2018) 253–262. https://doi.org/10.1016/j.actbio.2018.05.045.

[88]    L. Qian, S. Wang, S. Zhou, Y. Sun, H. Zhao, Influence of pia-arachnoid complex on the indentation response of porcine brain at different length scales, J Mech Behav Biomed Mater 127 (2022). https://doi.org/10.1016/j.jmbbm.2021.104925.

[89]    J. Malda, J.C. de Grauw, K.E.M. Benders, M.J.L. Kik, C.H.A. van de Lest, L.B. Creemers, W.J.A. Dhert, P.R. van Weeren, Of Mice, Men and Elephants: The Relation between Articular Cartilage Thickness and Body Mass, PLoS One 8 (2013). https://doi.org/10.1371/journal.pone.0057683.

[90]    A.M. Corson, J. Laws, A. Laws, J.C. Litten, I.J. Lean, L. Clarke, Percentile growth charts for biomedical studies using a porcine model, Animal 2 (2008) 1795–1801. https://doi.org/10.1017/S1751731108002966.

[91]    H. Ozawa, T. Matsumoto, T. Ohashi, M. Sato, S. Kokubun, Mechanical properties and function of the spinal pia mater, 2004.

[92]    P. Sudres, M. Evin, E. Wagnac, N. Bailly, L. Diotalevi, A. Melot, P.-J. Arnoux, Y. Petit, Tensile mechanical properties of the cervical, thoracic and lumbar porcine spinal meninges, J Mech Behav Biomed Mater 115 (2021). https://doi.org/https://doi.org/10.1016/j.jmbbm.2020.104280.

[93]    A.R. Tunturi, Elasticity of the spinal cord, pia, and dentieulate ligament in the dog, J Neurosurg 48 (1978) 975–979.

[94]    H. Kimpara, Y. Nakahira, M. Iwamoto, K. Miki, Investigation of Anteroposterior Head-Neck Responses during Severe Frontal Impacts Using a Brain-Spinal Cord Complex FE Model, Stapp Car Crash J 50 (2006) 509–544.

[95]    J.T. Maikos, R.A.I. Elias, D.I. Shreiber, Mechanical properties of dura mater from the rat brain and spinal cord, J Neurotrauma 25 (2008) 38–51. https://doi.org/10.1089/neu.2007.0348.

[96]    A. Tamura, W. Yano, D. Yoshimura, S. Nishikawa, Mechanical Characterization of Spinal Dura Mater Using a PD-Controlled Biaxial Tensile Tester, J Mech Med Biol 20 (2020). https://doi.org/10.1142/S0219519420500232.

[97]    D.R. Walsh, A.M. Ross, D.T. Newport, Z. Zhou, J. Kearns, C. Fearon, J. Lorigan, J.J.E. Mulvihill, Mechanical characterisation of the human dura mater, falx cerebri and superior sagittal sinus, Acta Biomater 134 (2021) 388–400. https://doi.org/10.1016/j.actbio.2021.07.043.

[98]    A. Tamura, S. Nishikawa, Effect of Anatomical Sites on the Mechanical Properties of Spinal Dura Subjected to Biaxial Stretching, J Eng Sci Med Diagn Ther 5 (2022). https://doi.org/10.1115/1.4053341.

[99]    A. Tamura, S. Nishikawa, DURAL MECHANICAL RESPONSES TO LOAD-CONTROLLED ASYMMETRIC BIAXIAL STRETCH, J Mech Med Biol 23 (2023). https://doi.org/10.1142/S021951942350077X.

[100]   A. Tamura, C. Sakaue, Effects of surface profile on porcine dural mechanical properties, ClinicalBiomechanics (2024).

[101]   J. Dobbing, The Later Growth of the Brain and its Vulnerability, Pediatrics 53 (1974). http://publications.aap.org/pediatrics/article-pdf/53/1/2/942205/2.pdf.

[102] J. Dobbing, The Later Development of the Brain and its Vulnerability, J Inherit Metab Dis 5 (1982) 88.

[103] J. Dickerson, J. Dobbing, Prenatal and postnatal growth and development of the central nervous system of the pig, Proceedings of the Royal Society 166 (1967) 384–395.

[104] K.L. Thibault, S.S. Margulies, Material Properties of the Developing Porcine Brain, in: International Research Council on Biomechanics of Injury, 1996: pp. 75–85.

[105] P. Saboori, Subarachnoid trabeculae, in: Cerebrospinal Fluid and Subarachnoid Space: Clinical Anatomy and Physiology: Volume 1, Elsevier, 2022: pp. 213–228. https://doi.org/10.1016/B978-0-12-819509-3.00006-7.

[106] P. Aimedieu, R. Grebe, Tensile strength of cranial pia mater: preliminary results, J Neurosurg (2004) 111–114.

# Supplemental Information for

## Mechanical Properties of the Meninges:

## Large Language Model Augmented Systematic Review

Brandon P. Chelstrom, Maciej P. Polak, Dane Morgan, Corinne R. Henak

## 1) Validation of the LLM-based data extraction

Precision and recall were evaluated using ten sample studies. Briefly, property triplets from ten studies were manually extracted and assumed to be the ground truth. The sample studies were selected to target a search with the material as the *dura mater* and property of elastic modulus and varied with respect to publishing year (1960 to 2023), authors, journals, study type, study length, and relevance. Property triplets from ChatExtract were then compared to the ground truth where equivalent property triplets were defined as having identical units, identical values, and property text which uniquely describe the property (e.g. elastic modulus and Young's modulus would be equivalent, but elastic modulus and shear modulus would not). Precision and recall were calculated [1]:

$$Precision = \frac{TruePositives}{TruePositives + FalsePositives}$$

$$Recall = \frac{TruePositives}{TruePositives + FalseNegatives}$$

To establish the effect of LLM on the results on the semi-automatic review, the modified ChatExtract algorithm was run 10 times on two different LLMs (GPT-3.5-turbo and GPT-4o) for both a specific property value (elastic modulus) and a more general property value (stiffness) (Fig. S1). For the specific property of elastic modulus, GPT-4o had perfect recall with moderate precision indicating all correct property triplets were extracted with additional irrelevant property triplets. The inverse was true for the general property of stiffness: GPT-4o had perfect precision with moderate recall suggesting that while some property triplets were missed, all extracted property triplets were valid. GPT-3.5-turbo performed relatively the same for the specific and general property with moderate recall and precision indicating some property triplets were not included while other irrelevant property triplets were included (Fig S1).

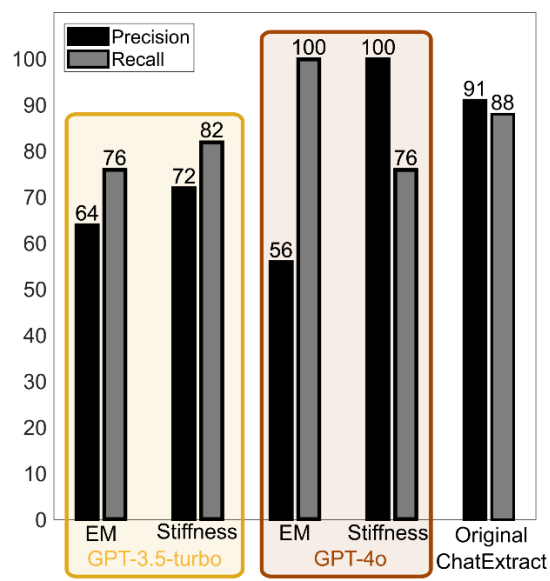

Fig S1: Precision and recall validation results from both GPT-3.5-turbo and GPT-4o for both specific (elastic modulus) and general (stiffness) properties compared to the original ChatExtract algorithm.

*2) Additional Figures and Tables*

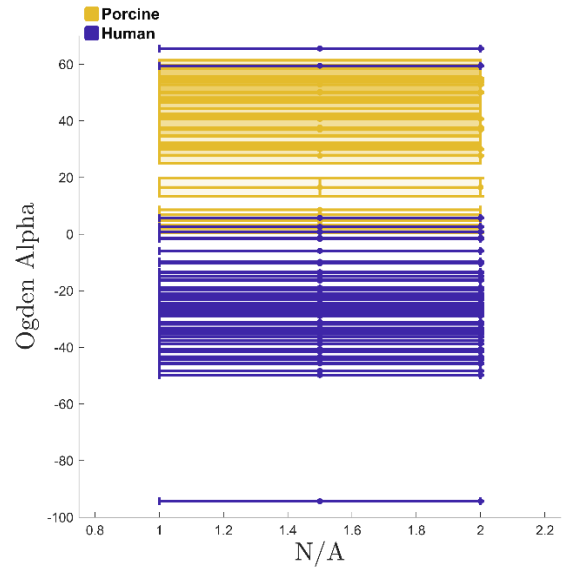

Fig S2: Distribution of Ogden exponentials for human (purple) and porcine (orange) data. All values were standardized to the uniaxial incompressible form of $W = \frac{\mu}{\alpha^2}(\lambda_1^\alpha + \lambda_2^\alpha + \lambda_3^\alpha)$, where $\mu$ is the stiffness, $\alpha$ is the exponential term, and $\lambda_i$ are the principal stretches.

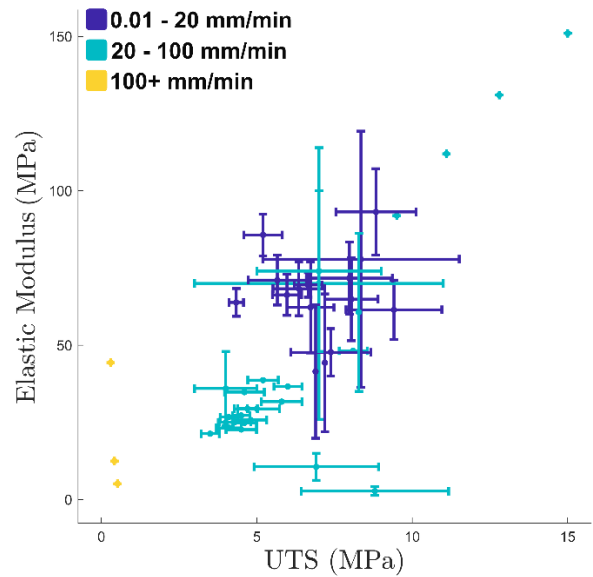

Fig S3: Ultimate tensile stress (UTS) versus elastic modulus for the *dura mater* at various loading rates. Plotted values are the average and standard deviation for various studies at various loading rates. Purple corresponds to slow loading rates ranging between 0.01 and 20 mm/min, green corresponds to moderate loading rates ranging between 20 and 100 mm/min, and yellow corresponds to fast loading rates above 100 mm/min.

Table S1: Relevant units for each property used for preprocessing of relevant sentences

| Property | Relevant Units |
|---|---|
| Elastic Modulus | Pa, kPa, MPa, GPa |
| Young's Modulus | Pa, kPa, MPa, GPa |
| Shear Modulus | Pa, kPa, MPa, GPa |
| Relaxation Time | s, ms, µs |
| Strain | - |
| Strain Rate | - |
| Rate Dependence | - |
| Stiffness | - |
| Nonlinear | - |
| Temperature | C, F, K |
| Linear Elastic | - |
| Viscoelastic | - |
| Pressure | - |
| Finite Element Parameter | - |
| Model | - |

Table S2: Summary of dura mater mechanical papers, papers are arranged chronologically. N/R indicates that data were note reported in the reference, but atmospheric test conditions are assumed to be open to room when not reported. Post Mortem Time (PMT), Standard Deviation (SD).

| Authors | DOI | Sample/ Cadaver Number | Cranial/ Spinal | Species | Location | Orientation | Treatment/ Study Purpose | Mean Age (SD) <range> [years] | Females /Males | PMT (SD) <range> [hours] | Sample Storage | Atmospheric Test Conditions | Sample Size (SD) <range> [mm] | Vascular/ Avascular | Loading Rate |
|---|---|---|---|---|---|---|---|---|---|---|---|---|---|---|---|
| (Galford and McElhaney, 1970) [2] | https://doi.org/10.1016/0021-9290(70)90007-2 | 11/2 | Cranial | Human/ Monkey | N/R | N/R | Dynamic tissue properties | N/R | N/R | <6-12> | Maintained in ringer solution | Kept Moist with ringer solution | Rectangular, 63.5 mm x 6.35 mm | N/R | N/R |
| (Melvin et al., 1970) [3] | https://www.jstor.org/stable/44723768 | 36/- | Cranial | Human | N/R | Longitudinal, Transverse, Diagonal | Tissue orientation | N/R | N/R | 3 | Stored in saline solution | Open to Room | Dog Bone, Gage: 19.05 mm x 6.35 mm | N/R | 0.061 s$^{-1}$ |
| (R. M. Kenedi, 1973) [4] | 839107099 | 123/30 | Cranial | Human | Parietal, Temporal, Frontal | N/R | Influence of strain rate | N/R | N/R | 10 | Stored in saline solution | Open to Room | Dog Bone, 19.05 mm x 6.35 mm | N/R | 0.0666, 0.666, 6.66 s$^{-1}$ |
| (Van Noort et al., 1981a) [5] | https://doi.org/10.1016/0142-9612(81)90086-7 | 24/16 | Cranial | Human | N/R | N/R | Half of samples preserved in glycerol | 56 (16.5) <20 - 77> | N/R | <12-17> | Refrigerated saline | Open to Room | Rectangular, 40 mm x 5 mm | Avascular | 50 mm/min |
| (Van Noort et al., 1981b) [6] | https://doi.org/10.1088/0143-0815/2/3/003 | 260/20 | Cranial | Human | N/R | N/R | Some samples preserved in saline or glycerol | <28 - 88> | Both | <12-17> | Refrigerated saline | Open to Room | Rectangular, 60 mm x 5 mm, (Guage 40 mm x 5 mm) | Avascular | 50 mm/min |
| (Zyablov et al., 1982) [7] | N/R | 226/150 | Cranial | Human | N/R | Longitudinal, transverse | Age effects on mechanics and physiology | <-0.917 - 90> | N/R | N/R | N/R | Open to Room | Rectangular, 100 mm x 30 mm | N/R | 0.016 N/min |

| Reference | DOI | Ratio | Type | Species | Region | Orientation | Purpose | Value | Ratio2 | | Storage | Environment | Sample Shape | | Rate |
|---|---|---|---|---|---|---|---|---|---|---|---|---|---|---|---|
| (Kriewall et al., 1983) [8] | N/R | 9/9 | Cranial | Human | Parietal | Left, right | Mechanical behavior of fetal tissue | -0.136 (0.1) <-0.42 - 0> | 5/4 | N/R | Refrigerated at 5 or -10°C | Open to room (23C) but maintained in saline solution | Circular, 60 mm diameter | N/R | 0.007 MPa/s |
| (McGarvey et al., 1984) [9] | https://doi.org/10.1016/0142-9612(84)90011-5 | 54/13 | Cranial | Human | By sagittal sinus | Longitudinal, transverse | Half of samples preserved in glycerol | 52 <17-72> | 3/10 | 20 | Maintained in Hanks balance solution (pH 7.4, osmolarity 310 mOs) | Maintained in Hanks solution at 37°C | Rectangular, 10.1 (0.2) mm x 7 mm | N/R | 5, 50, 100, 500 %/min |
| (Tencer et al., 1985) [10] | https://doi.org/10.1097/00007632-198510000-00009 | 40/5 | Spinal | Human | Posterior/Anterior of lumbar, low thoracic, high thoracic, cervical | Longitudinal | Mechanical testing following spinal extension/flexion | <N/R - 65> | N/R | N/R | Wrapped in wet towels at -20°C | Open to Room | Dog Bone, 38.1 mm x 2.54 mm | N/R | N/R |
| (Bylski et al., 1986) [11] | https://doi.org/10.1016/0021-9290(86)90105-3 | 11/7 | Cranial | Human | Parietal/Frontal | Left, right | Biaxial mechanical behavior of fetal tissue | -0.11 (0.096) <-0.286 - 0> | 3/4 | N/R | Refridated at 5 or frozen at -10°C | Open to room but maintained in saline solution | Circular, 40 mm diameter | N/R | 0.02 mm/s |
| (Patin et al., 1993) [12] | https://doi.org/10.1213/00000539-199303000-00014 | 9/9 | Spinal | Human/dog | Dorsal thoracolumbar | Longitudinal, transverse | Isotropy of spinal tissue | 43.36 (29.22) <0.041-81> | 3/6 | N/R | Refrigerated in saline | Open to Room | Rectangular, 15 mm x 10 mm | N/R | 100 mm/min |
| (Wolfinbarger et al., 1994) [13] | https://doi.org/10.1002/jab.770050313 | 95/10 | Cranial | Human | Noted, but tested random | Noted, but tested random | Tissue freeze-dried prior to testing | 40.3 (3.8) <17 - 51> | 1/7 | <2-3> | Freeze-dried, then stored on ice | Open to Room | Rectangular, 40 mm x 10 mm | N/R | 10 mm/min |
| (Mickley et al., 1994) [14] | https://doi.org/10.1016/0031-9384(94)90384-0 | 97/- | Cranial | Rat | N/R | N/R | Thermal dose effect on surface temperature | N/R | N/R | 0 | N/R | In Situ | N/R | N/R | N/R |
| (E. Zarzur, 1996) [15] | https://doi.org/10.1590/S0004-282X1996000300015 | 6/3 | Spinal | Human | Posterior lumbar | Longitudinal, transverse | Samples preserved in formalin | 56.33 (17.55) <38 - 73> | -/3 | 72 | Stored in formalin | Open to Room | Rectangular, 20 mm x 20 mm | N/R | 20 mm/min |
| (Yamada et al., 1997) [16] | https://doi.org/10.3171/jns.1997.86.6.1012 | 15/15 | Cranial | Human | N/R | N/R | Compare natural and synthetic mechanical behavior | 30.7 (20.7) <3 62> | N/R | N/R | N/R | Open to room, sprayed with saline | Rectangular, 40 mm x 5 mm, (Gage 10 mm x 5 mm) | N/R | 50 mm/min |

| | | | | | | | | | | | | | | | | |
|---|---|---|---|---|---|---|---|---|---|---|---|---|---|---|---|---|
| (Sacks et al., 1998) [17] | https://doi.org/10.1115/1.2798027 | 90/5 | Cranial | Human | Noted, but tested random | Parallel, perpendicular to fibers | Isotropy of cranial tissue, half subjected to allograft preparation | 54 (22) | N/R | N/R | | Frozen with saline in liquid nitrogen vapor | Submerged in room temperature saline | Rectangular, 20 mm x 4 mm | N/R | 50 mm/min |
| (Runza et al., 1999) [18] | https://doi.org/10.1213/00000539-199906000-00022 | /6 (Human), /2 (Bovine) | Spinal | Human/Bovine | Dorsalum bar (T12-L4/L5) | Longitudinal, transverse | PMT and storage temperature | 58.5 <39 - 86> | 3/3 | | 2, 24, 120 | Fresh, -4°C for 24 h, -4°C for 120 h, 4°C for 96 h | Open to room (20°C, 61% humidity) | Dog Bone, Gage: 22 mm x 4 mm | N/R | 10 mm/min |
| (Wang et al., 2001) [19] | https://doi.org/10.1016/S0166-4328(00)00395-8 | 3/3 | Cranial | Rat | Occipital | N/R | Dura cold probe temperature dispersion | N/R | N/R | | 0 | N/R | In Situ | N/R | N/R | 3 min |
| (Wilcox et al., 2003) [20] | https://doi.org/10.1007/s10776-003-0644-9 | 9/- | Spinal | Bovine | N/R | Longitudinal, transverse | Time-dependent and directional properties of spinal tissue | N/R | N/R | | 2 | Frozen and thawed prior to testing | Submerged in saline (37°C) | Rectangular, 6-8 mm x 30-40 mm | N/R | 0.03 - 0.13 s$^{-1}$ |
| (Bashkatov et al., 2003) [21] | https://doi.org/10.1016/S0006-3495(03)74750-X | N/R | Cranial | Human | N/R | N/R | Glucose and mannitol diffusion behavior | N/R | N/R | | 24 | Stored at -12°C | Open to room (~20°C) | Rectangular, 10 mm x 10 mm | N/R | N/R |
| (Maikos et al., 2008) [22] | https://doi.org/10.1089/neu.2007.0348 | 23/- (Spinal), 16/- (Cranial) | Cranial/Spinal | Rat | Noted, but tested random | Longitudinal (Spinal) | Determine mechanical properties of rat dural tissue | 0.21 (0.0136) | N/R | | 2 | Stored in PBS | Open to room, but hydrated via ultrasonic humidifier | Rectangular, 12 mm x 1.5-2.5 mm | N/R | 0.0014 or 19.4 s$^{-1}$ |
| (Persson et al., 2010) [23] | https://doi.org/10.1007/s10439-010-9924-6 | 38/20 | Spinal | Bovine | Lumbar | Longitudinal, transverse | Evaluate nonlinear behavior of spinal tissue | <1.33 - 2> | N/R | | N/R | Saline wrapped at -20°C | Open to room, but hydrated with saline solution | Dog Bone, 40 mm x 10 mm (Gage 10 mm x 4 mm) | N/R | 15, 150, 1500 mm/min (0.01, 0.1, 1.0 s$^{-1}$) |
| (Mazgajczyk et al., 2012) [24] | N/R | 216/9 | Spinal | Porcine | Cervical (C1-C7), ventral/dorsal | Longitudinal, transverse | Location and direction dependence of dura mater mechanical properties | N/R | N/R | | 24 | Frozen, then prepped in 0.9% saline at 4°C | Open to room (24°C) | Rectangular, 8.1 (0.4) mm x 5.3 (0.3) mm | N/R | 2 mm/min |
| (Shetye et al., 2014) [25] | https://doi.org/10.1016/j.jmbbm.2014.02.014 | 12/4 | Spinal | Ovine | Cervical (C1-C6) | Longitudinal, transverse | Biaxial characterization of fiber alignment | N/R | N/R | N/R | N/R | N/R | Open to Room | Rectangular 25 mm x 25 mm | N/R | N/R |

| Reference | DOI | n/N | Type | Species | Location | Direction | Purpose | | | | Storage | Temperature | Specimen | | Rate |
|---|---|---|---|---|---|---|---|---|---|---|---|---|---|---|---|
| (Ramo et al., 2018) [26] | https://doi.org/10.1115/1.4038261 | 45/4 | Spinal | Ovine | Cervical (C2-C7) | Longitudinal | Characterize development of damage accumulation | N/R | N/R | N/R | Saline wrapped at -20°C | Saline bath at room temperature | Rectangular 20.81 (3.17) mm x 4.29 (0.9) mm | N/R | 0.6, 60, 360 mm/min (0.0005, 0.051, 0.284 s$^{-1}$) |
| (Walsh et al., 2018) [27] | https://doi.org/10.1016/j.actbio.2018.09.004 | 171/9 | Cranial | Porcine | Frontal, temporal, parietal, occipital, sagittal sinus | Medial/ Lateral, Rostral/ Caudal | Biological, geometric, and location dependence of spinal tissue | 0.583 | N/R | 48 | Stored in PBS at 4°C | Open to room, but warmed to 37°C in PBS bath prior to testing | Rectangular, 20 mm x 2 mm (Frontal, occipital, parietal, temporal), 30 mm x 4 mm (Sagittal Sinus) | N/R | 1 s$^{-1}$ |
| (Kegel et al., 2018) [28] | https://doi.org/10.1016/j.jmbbm.2017.12.023 | 53/5 | Cranial | Human | Posterior, middle, anterior | Left, right | Extensive nonlinear mechanical characterization of spinal tissue | <68 - 88> | 1/4 | <96 - 144> | Dry-frozen at -18°C, thawed to 4°C | Open to Room | Rectangular, 10 mm x 10 mm | N/R | 3.6 mm/min (1 s$^{-1}$) |
| (Aydin et al., 2019) [29] | https://doi.org/10.3340/jkns.2018.0130 | 28/7 | Cranial | Human | Frontal | Left, right | Determine change of mechanical properties due to needle punctures | 44.57 (11.58) <31 - 62> | 4/3 | 30 | Stored at -20°C, thawed at 4°C for 24 hrs, thawed at 20-25°C for 6 hours | Open to Room | Dog Bone, Gage 20 mm x 5 mm | N/R | 10 m/min |
| (Zwirner et al., 2019a) [30] | https://doi.org/10.1038/s415 98-019-52836-9 | 126/75 | Cranial | Human | Temporal | Left, right | Influence of age, sex, and post mortem on mechanical properties | 50 (23) <2 - 94> | 25/48 | 75 (30) <11- 146> | Cadavers stored at 4°C, samples frozen at -80°C | Open to Room | Dog Bone, Gage: 50 mm x 10 mm | N/R | 20 mm/min |
| (Chengwei m et al., 2019) [31] | https://doi.org/http://dx.doi.org/10.7507/1002-1892.201807085 | -/5 | Spinal | Ovine | Cervical (C6, C7), thoracic (T11, T12), lumbar (L4, L5) | Longitudinal, transverse | Investigate directional behavior of spinal tissue | 1 | -/5 | N/R | Placed in saline | Open to room (25°C and 85% humidity) | Rectangular, 20 mm x 5 mm | N/R | 0.96 mm/min |
| (Zwirner et al., 2019b) [32] | https://doi.org/10.1016/j.jmbbm.2019.04.035 | 36/18 | Cranial | Human | N/R | N/R | Comparison between acellular and native tissue | 48 <12-83> | 6/12 | 71 (28) <14- 121> | Samples frozen at -80°C | Open to Room | Dog Bone, Gage: 20 mm x 5 mm | N/R | 20 mm/min |
| (Lipovka et al., 2020) [33] | https://doi.org/10.1088/1742-6596/1661/1/012062 | 3/3 | Cranial | Human | N/R | N/R | Tissue harvested during microsurgical treatment method | N/R | N/R | <12 - 48> | Preserved in 0.9% saline at 2-5°C | Open to Room | Rectangular | N/R | 2 mm/min |

| Reference | DOI | | Cranial/Spinal | Species | Region | Direction | Purpose | | | | Storage | Testing Environment | Sample | | Rate |
|---|---|---|---|---|---|---|---|---|---|---|---|---|---|---|---|
| (Pierrat et al., 2020) [34] | https://doi.org/10.3389/fbioe.2020.00801 | -/11 | Cranial | Porcine | Parietal | Longitudinal, transverse | Nonlinear characterization for finite element models | 0.432 (0.01) <0.423 - 0.442> | Mixed | <4 - 24> | Refrigerated in damp tissue | Open to room, but kept in saline solution prior to testing | Dog Bone, 63.5 mm x 3.2 mm | N/R | 50 mm/min (0.01 s$^{-1}$) |
| (Tamura et al., 2020) [35] | https://dx.doi.org/10.1142/S02195194 20500232 | 18/6 | Spinal | Porcine | Posterior thoracic and lumbar | Longitudinal, transverse | Verification of custom biaxial test device | N/R | N/R | 24 | Stored in saline at 4°C | Submerged in saline (37°C) | Rectangular, 15 mm x 15 mm | N/R | 21 mm/min |
| (Walsh et al., 2021) [36] | https://doi.org/10.1016/j.actbio.2021.07.043 | 82/7 | Cranial | Human | Frontal, parietal, temporal, occipital, falx cerebri, superior sagittal sinus | Medial/ Lateral, Anterior/ Posterior | Biaxial characterization of cranial tissue | <30-90> | 1/6 | N/R | Stored at -80°C with DMEM (1.8M) and sucrose (0.1M), warmed to 37°C in water bath, submerged in PBS at 4°C | Submerged in PBS bath (37°C) | Rectangular, 10 mm x 10 mm or 5 mm x 5 mm (Biaxial samples not sectioned) | N/R | N/R |
| (Su et al., 2022) [37] | https://doi.org/10.1016/j.cjmeam.2022.100026 | 5/- | Cranial | Porcine | N/R | N/R | Fixed in 3.7% paraformaldehyde for 24 h at 4°C, dehydrated with alcohol aqueous solutions, and freeze-dried | N/R | N/R | 24 | Fixed in 3.7% paraformaldehyde for 24 h at 4°C, dehydrated with alcohol aqueous solutions, and freeze-dried | Open to Room | Rectangular, 50 mm x 6 mm | N/R | 2 mm/min |
| (Pearcy et al., 2022) [38] | https://doi.org/10.1016/j.jmbbm.2021.104866 | -/32 | Cranial | Human | Temporal and parietal | Parallel and perpendicular to principal artery | Samples crosado-embalmed prior to testing | 81 (9) <58 - 95> | 16/16 | N/R | Samples preserved with crosado mix | Open to room (22°C) | Dog Bone, Gage 10 mm x 5 mm | Longitudinal and transverse to vasculature | 20 mm/min |
| (Cavdar et al., 2022) [39] | https://doi.org/10.1016/j.wneu.2021.12.029 | -/30 | Cranial | Human | Frontal, parietal, temporal, occipital | Longitudinal (sagittal), transverse | Compare cortical thickness and histological measures | Female: 40.57 <27 - 63>, Males: 46.25 <32 58> | 14/16 | N/R | N/R | Open to Room | Rectangular, 30 mm x 10 mm | N/R | 30 mm/min |
| (Cavelier et al., 2022) [40] | https://doi.org/10.1016/j.jmbbm.2021.105056 | 88/- (Spinal), 18/- (Cranial) | Spinal/ Cranial | Porcine (landrace) | Dorsal and ventral cervical, thoracic, lumbar (Spinal), Frontal (Cranial) | Longitudinal, transverse | Compare directionality of cranial and spinal tissues | <0.25 - 0.33> | N/R | stored for <6 months, then 15 | Wrapped in saline-soaked gauze at -20°C, defrosted for 15 h at 4°C, prepped in PBS | Continuous PBS flow at room temperature | Spinal Samples: Rectangular, 36 (2) mm x 5 (1) mm, Cranial Samples: Rectangular, 40 mm x 40 mm, Gage 16 | N/R | 15 mm/min |
| | | | | | | | | | | | | | (2) mm x 5 (1) mm | | |
| (Evin et al., 2022) [41] | https://doi.org/10.1016/j.actbio.2021.11.028 | 31/8 | Spinal | Porcine (landrace) | Cervical (C0-C7), Thoracic (T1-T15), Lumbar (L1-L5) | Longitudinal, transverse | Characterize spinal dura and pia for consitutive model development | <0.25 - 0.33> | N/R | 12 | Prepped in saline solution | Open to room, but hydrated with water droplets | Rectangular, 7 mm x 7 mm | N/R | 3 mm/min |
| (Tamura et al., 2022) [42] | https://doi.org/10.1115/1.4053341 | 91/18 | Spinal | Porcine | Dorsal and ventral cervical (C1-T1), thoracic (T2-T15), lumbar (L1-L5) | Longitudinal, transverse | Determine anatomical variation of spinal tissue | N/R | N/R | N/R | Stored in saline solution at 4°C | Floating in PBS bath at 37°C | Rectangular, 12 mm x 12 mm | N/R | 12 mm/min |
| (Niestrawska et al., 2023) [43] | https://doi.org/10.1016/j.actbio.2023.08.024 | 90/6 | Cranial | Human | Left ventral medial, lateral, medial, dorsal lateral, dorsal medial | Relative longitudinal, transverse | Link cranial microstructure heterogeneity to mechanical properties | 78 (6) <70 - 87> | 1/5 | 48 | Frozen at -80°C, cooled to 4°C over 3 h | Submerged in PBS (pH = 7.4) bath (37°C) | Dog bone, Gage 20 mm x 5 mm | N/R | 2 mm/min |
| (Sharma et al., 2023a) [44] | https://doi.org/10.1016/j.bea.2023.100108 | 8/6 | Cranial | Porcine (Yorkshire) | N/R | N/R | Mechanical compare native and decellularized tissue | <0.5 - 0.67> | -/6 | 2 (native), 76 (decellularized) | Freeze-dried in liquid nitrogen | Open to room, but sprayed with PBS | Circular, 14 mm diameter | N/R | 0.6 mm/min |
| (Sharma et al., 2023b) [45] | https://doi.org/10.1016/j.engreg.2023.02.004 | 36/12 (uniaxial tension), 60/2 (microindentation) | Cranial | Porcine (Yorkshire) | Parietal (uniaxial tension), anterior, middle, posterior, left, and right (microindentation) | Longitudinal (anterior-posterior) | Mechanical compare native and decellularized tissue | <0.5 - 0.67> | -/12 | 3 | Samples kept in 1xPBS | Open to Room | Dog Bone, 40 mm x 10 mm (Gage 10 mm x 4 mm) | N/R | 0.01 s$^{-1}$ (uniaxial tension), 0.6 mm/min (microindentation) |

| Authors | DOI | Sample/Cadaver Number | Cranial/Spinal | Species | Location | Orientation | Treatment/Study Purpose | Mean Age (SD) <range> [years] | Females/Males | PMT (SD) [hours] | Sample Storage | Atmospheric Test Conditions | Sample Size (SD) <range> [mm] | Vascular/Avascular | Loading Rate |
|---|---|---|---|---|---|---|---|---|---|---|---|---|---|---|---|
| (Zwirner et al., 2023) [46] | https://doi.org/10.1016/j.jmbbm.2022.105617 | 30/4 | Cranial | Human | Occipital | Longitudinal | Evaluate rate dependence of dura mater, tissue fixed in PEG-Tris (pH = 7.4) at 4°C for 24 hrs | 76 (6) | 1/3 | 48 | Cadavers stored at 4°C, samples frozen at -80°C | Open to room (22°C) | Dog bone, Gage: 20 mm x 5 mm | N/R | 18000, 30000, 42000 mm/min (15, 25, 35 s⁻¹) |
| (A. Tamura et al., 2023) [47] | https://doi.org/10.1142/S021951942350077X | 26/10 | Spinal | Porcine | Lumbar | Longitudinal, transverse | Compared spinal samples with and without elastase treatment | N/R | N/R | 24 | Stored in saline solution at 4°C | Open to Room | Rectangular, 12 mm x 12 mm | N/R | 12 mm/min |
| (Tamura and Sakaue, 2024) [48] | https://doi.org/10.1016/j.clinbiomech.2024.106189 | 55/10 | Spinal | Porcine | Cervical, lumbar | Longitudinal, transverse | Determine effect of surface roughness on mechanics | N/R | N/R | 48 | mounted on agar gel (1:25 mixing weight), submerged in PBS at 4°C | Sample covered with windshield at room temperature (~22°C) | Rectangular, 12 mm x 12 mm | N/R | 0.9 mm/min |

Table S3: Summary of arachnoid mater mechanical papers and pia-arachnoid complex (PAC) papers, papers are arranged chronologically. N/R indicates that data were note reported in the reference, but atmospheric test conditions are assumed to be open to room when not reported.

| Authors | DOI | Sample/Cadaver Number | Cranial/Spinal | Species | Location | Orientation | Treatment/Study Purpose | Mean Age (SD) <range> [years] | Females/Males | PMT (SD) [hours] | Sample Storage | Atmospheric Test Conditions | Sample Size (SD) <range> [mm] | Vascular/Avascular | Loading Rate |
|---|---|---|---|---|---|---|---|---|---|---|---|---|---|---|---|
| (Jin et al., 2006) [49] | https://doi.org/10.4271/2006-22-0025 | 40/10 | Cranial | Bovine | Dorsal | Longitudinal (sagittal), transverse (coronal) | Mechanically characterize the cranial PAC in tension | <0.327 - 0.385> | N/R | 48 | Stored in artificial cerebral spinal fluid | Open to room, but sprayed with artificial cerebral spinal fluid | Rectangular, 65 mm x 30 mm, (Gage 20 mm x 25.6 mm) | N/R | 60, 600, 6000, 120000 mm/min (0.05, 0.5, 5, 100 s⁻¹) |
| (Jin et al., 2007) [50] | https://doi.org/10.4271/2007-22-0004 | 40/- | Cranial | Bovine | Frontal, occipital, parietal | N/R | Mechanically characterize the cranial PAC in traction | <0.327 - 0.385> | N/R | 48 | Soaked in artificial cerebral spinal fluid | Open to room, but sprayed with artificial cerebral spinal fluid | Rectangular, 12.7 mm x 12.7 mm | N/A | 0.516, 2.82, 28.8, 165 mm/min (0.36, 2.0, 20.5, 116.3 s⁻¹) |

| Reference | DOI | | Type | Species | Region | | Purpose | | | | Condition | Environment | Geometry | | Rate |
|---|---|---|---|---|---|---|---|---|---|---|---|---|---|---|---|
| (Jin et al., 2011) [51] | https://doi.org/10.1016/j.jbiomech.2010.09.035 | 43/4 | Cranial | Bovine | Frontal, occipital, parietal | N/R | Mechanically characterize the cranial PAC in shear | <0.327 - 0.385> | N/R | 48 | Soaked in artificial cerebral spinal fluid | Open to room (20°C), but sprayed with artificial cerebral spinal fluid | Rectangular, 12.7 mm x 12.7 mm | N/R | 1.2, 10.2, 102 mm/min (0.84, 7.3, 72 s$^{-1}$) |
| (Jin et al., 2014) [52] | https://doi.org/10.1007/s10439-013-0948-6 | N/R | Cranial | Bovine | N/R | N/R | Develop constitutive equations for the macroscopic cranial PAC | <0.327 - 0.385> | N/R | 48 | Soaked in artificial cerebral spinal fluid | Open to room (20°C), but sprayed with artificial cerebral spinal fluid | Rectangular, 12.7 mm x 12.7 mm (normal traction and shear), Rectangular, Gauge 20 mm x 25.6 mm (Uniaxial tension) | N/R | 0.05, 0.5, 5, 100 s$^{-1}$ |
| (Ramo et al., 2018) [53] | https://doi.org/10.1016/j.actbio.2018.05.045 | 8/8 | Spinal | Ovine | Cervical (C0-C7) | Longitudinal | PAC tested following full cord testing | <4 - N/R> | 8/- | 5 | Soaked in saline solution in airtight container | Open to room | Rectangular, 60 mm x 10 mm | N/R | 0.016, 0.16, 1.6, 16.0 s$^{-1}$ |
| (Fabris et al., 2019) [54] | https://doi.org/10.1016/j.actbio.2019.09.044 | 18/12 | Cranial | Rat (Sprague-Dawley) | Cortical, cerebellar | N/R | Determine effects of vasculature on PAC properties | <0.154 - 0.231> | -/12 | 0.33 | Soaked in PBS | Sample immersed in PBS | Rectangular, grid of 99 points in a 90 μm x 90 μm area | Noted by AFM and immunofluorescence | 0.6 - 3 mm/min (0.33 - 1.66 s$^{-1}$) |
| (Benko et al., 2021) [55] | https://doi.org/10.1016/j.jmbbm.2021.104579 | 109/5 | Cranial | Human | Superior and Inferior regions of the frontal, occipital, parietal | N/R | Determine traction modulus of individual arachnoid trabeculae | 60 (19.2) <32 - 81> | 2/3 | 56 | Frozen and defrosted over 2 days to room temperature, Inflated with Saline | In Situ | Tissue not sectioned, scan volume of 5 mm x 5 mm x 1.64 mm | N/R | 0.00159 MPa/min |
| (Qian et al., 2022) [56] | https://doi.org/10.1016/j.jmbbm.2021.104925 | -/32 | Cranial | Porcine | N/R | Left/Right pooled | Determine mechanical behavior of PAC across length scales | 0.5 | N/R | 6 | Cortex submerged in artificial cerebral spinal fluid | Sample immersed in artificial cerebral spinal fluid at room temperature | N/R | N/R | 48 and 96 mm/min (0.64 and 1.27 s$^{-1}$) |

Table S4: Summary of pia mater mechanical papers and, papers are arranged chronologically. N/R indicates that data were note reported in the reference, but atmospheric test conditions are assumed to be open to room when not reported.

| Authors | DOI | Sample/ Cadaver Number | Cranial/ Spinal | Species | Location | Orientation | Treatment/ Study Purpose | Mean Age (SE) <range> [years] | Females/ Males | PMT (SE) <range> [hours] | Sample Storage | Atmospheric Test Conditions | Sample Size (SE) <range> [mm] | Vascular/ Avascular | Loading Rate |
|---|---|---|---|---|---|---|---|---|---|---|---|---|---|---|---|
| (A. Tunturi, 1978) [57] | https://doi.org/10.3171/jns.1978.48.6.09 75 | N/R | Spinal | Dog | Thoracic (T9 - T11) | Longitudinal | Determine static mechanical properties of the spinal cord | N/R | N/R | 1 | Sprayed with ringer solution | Sprayed with ringer solution | Rectangular, 40 mm x - mm | N/R | 0.049 N/min |
| (Ozawa, 2004) [58] | https://doi.org/10.3171/spi.2004.1.1.012 2 | 18/9 | Spinal | Rabbit | Cervical (C5 - C6) | Transverse | Determine influence of pia mater spine mechanical properties | N/R | N/R | 0.5 | Stored in saline solution at room temperature (16°C) | Open to Room | Whole spinal column, 10 mm x 3 mm | N/R | 0.012 N/min |
| (Aimedieu & Grebe, 2004) [59] | https://doi.org/10.3171/jns.2004.100.1.0 111 | 9/6 | Cranial | Bovine | Posterior | Right hemisphere | Determine the elastic properties of cranial pia mater | 0.865 (0.54) | N/R | <6 - 16> | Cortex submerged in 0.9% saline solution at 4°C | Submerged in water bath (37°C) | Rectangular, 95 mm x 35 mm | N/R | 0.5 mm/min |
| (Kimpara et al., 2006) [60] | https://doi.org/10.4271/2006-22-0019 | 52/6 | Spinal | Porcine | By denticulate ligament, posterior median septum, posterolateral sulcus | N/R | Determine localized mechanical properties of the spinal pia mater | N/R | N/R | 5 | Open to room (24°C) at >50% humidity | Open to room (24°C) at >50% humidity | Rectangular, 20 mm x 1 mm | N/R | 0.005, 0.05, 0.5 s⁻¹ |
| (Li et al., 2020) [61] | https://doi.org/10.1016/j.clinbiomech.2020.10510 8 | 48/20 | Cranial | Ovine | Left/Right Hemisphere sagittal sinus | N/R | Determine viscoelastic properties of cranial pia mater at different loading rates | <0.461 - 0.538> | N/R | 6 | Cortex transferred in 4°C, and prepped at room temperature (22°C), maintained in 0.9% saline solution | Open to room (22°C) | Rectangular, 20 mm x 3-5 mm (Gage 10 mm x 4 mm) | N/R | 1, 5, 10, 30, 100, 300, 500 mm/min (0.00167, 0.0083, 0.0167, 0.083, 0.083, 0.16, 0.5, 0.83 s⁻¹) |
| (Sudres et al., 2021) [62] | https://doi.org/10.1016/j.jmbbm.2020.104280 | 47/6 | Spinal | Porcine (Yorkshire x Landrace) | Cervical (C1 - T1), thoracic (T1 - T4), lumbar (T14 - L6) | Longitudinal | Mechanical characterization of the pia mater | <0.167 - 0.250> | 2/4 | 12.5 | Stored in PBS at 4°C | Open to Room (20°C) | Rectangular, 10.35 (1.98) mm x 9.82 (3.31) mm | N/R | 12 mm/min |
| (Evin et al., 2022) [41] | https://doi.org/10.1016/j.actbio.2021.11.028 | 31/8 | Spinal | Porcine (landrace) | Cervical (C0-C7), Thoracic (T1-T15), Lumbar (L1-L5) | Longitudinal, transverse | Characterize spinal dura and pia for constitutive model development | <0.25 - 0.33> | N/R | 12 | Prepped in saline solution | Open to room, but hydrated with water droplets | Rectangular, 7 mm x 7 mm | N/R | 3 mm/min |

### References

[1] M.P. Polak, D. Morgan, Extracting Accurate Materials Data from Research Papers with Conversational Language Models and Prompt Engineering, (2023). http://arxiv.org/abs/2303.05352.

[2] J.E. Galford, J.H. McElhaney, A viscoelastic study of scalp, brain, and dura, J Biomech 3 (1970) 211–221. https://doi.org/10.1016/0021-9290(70)90007-2.

[3] J.W. Melvin, J.H. Mcelhaney, V.L. Roberts, Development of a Mechanical Model of the Human Head — Determination of Tissue Properties and Synthetic Substitute Materials, SAE Transactions 79 (1970) 2685–2694. https://about.jstor.org/terms.

[4] R.M.. Kenedi, Perspectives in Biomedical Engineering, in: R.M.. Kenedi (Ed.), Biological Engineering Society, University Park Press, 1973: pp. 215–222.

[5] R. Van Noort, M.M. Black, T.R.P. Martin, S. Meanley, A study of the uniaxial mechanical properties of human dura mater preserved in glycerol, Biomaterials 2 (1981) 41–45.

[6] R. Van Noort, T.R.P. Martin, M.M. Black, A.T. Barker, C.G. Montero, The mechanical properties of human dura mater and the effects of storage media, Clin. Phys. Physiol. Meas 2 (1981) 197–203. https://doi.org/https://doi.org/10.1088/0143-0815/2/3/003.

[7] V.I. Zyablov, Y.N. Shapovalov, K.D. Toskin, V. V Tkach, V. V Zhebrovsky, L.S. Georgievskaya, Structure and Physical-Mechanical Properties of the Human Dural Mind in the Age Aspect, Histology and Embryology 132 (1982) 29–35.

[8] T.J. Kriewall, N. Akkas, D.I. Bylski, B. Program, J.W. Melvin, Mechanical Behavior of Fetal Dura Mater Under Large Axisymmetric Inflation, 1983. http://asmedigitalcollection.asme.org/biomechanical/article-pdf/105/1/71/5481347/71_1.pdf.

[9] K. Mcgarvey A., J. Micheal Lee, D. Boughner R., Mechanical Suitability of Glycerol-Preserved Human Dura Mater for Construction of Prosthetic Cardiac Valves, Biomaterials 5 (1984) 109–117. https://doi.org/10.1016/0142-9612(84)90011-5.

[10] A.F. Tencer, B.L. Allen, R.L. Ferguson, A Biomechanical Study of Thoracolumbar Spine Fractures with Bone in the Canal. Part III. Mechanical Properties of the Dura and Its Tethering Ligaments, Spine (Phila Pa 1976) 10 (1985) 741–747.

[11] D.I. Bylski, T.J. Kriewall, N. Akkas, J.W. Melvin, Mechanical Behavior of Fetal Dura Mater Under Large Deformation Biaxial Tension, J Biomech 19 (1986) 19–26. https://doi.org/https://doi.org/10.1016/0021-9290(86)90105-3.

[12] D.J. Patin, E.C. Eckstein, K. Harum, V.S. Pallares, Anatomic and Biomechanical Properties of Human Lumbar Dura Mater, Anesth Analg 76 (1993) 535–540. https://doi.org/10.1213/00000539-199303000-00014.

[13] L. Wolfinbarger, Y. Zhang, B.L.T. Adam, D. Homsi, K. Gates, V. Sutherland, Biomechanical aspects on rehydrated freeze-dried human allograft dura mater tissues, Journal of Applied Biomaterials 5 (1994) 265–270. https://doi.org/10.1002/jab.770050313.

[14] A. Mickley G., B. Cobb L., P. Mason A., S. Farrell, Disruption of a Putative Working Memory Task and Selective Expression of Brain c-fos Following Microwave-Induced Hyperthermia, Physiol Behav 44 (1994) 1029–1038.

[15] E. Zarzur, Mechanical Properties of the Human Lumbar Dura Mater, Arq Neuropsiquiatr 54 (1996) 455–460.

[16] K. Yamada, S. Miyamoto, I. Nagata, H. Kikuchi, Y. Ikada, H. Iwata, K. Yamamoto, Development of a dural substitute from synthetic bioabsorbable polymers, J Neurosurg 86 (1997) 1012–1017.

[17] M.S. Sacks, J. Hamann, S.E. Otano-Lata, T.I. Malinin, Local Mechanical Anisotropy in Human Cranial Dura Mater Allografts, J Biomech Eng 120 (1998) 541–544. http://asmedigitalcollection.asme.org/biomechanical/article-pdf/120/4/541/5767037/541_1.pdf.

[18] M. Runza, R. Pietrabissa, S. Mantero, A. Albani, V. Quaglini, R. Contro, Lumbar Dura Mater Biomechanics: Experimental Characterization and Scanning Electron Microscopy Observations, Anesth Analg 88 (1999) 1317–1321.

[19] W. Yuan, C. Kathleen C., The role of the dura in conditioned taste avoidance induced by cooling the area postrema of male rats, Behavioural Brain Research 122 (2001) 113–129. https://doi.org/https://doi.org/10.1016/S0166-4328(00)00395-8.

[20] R.K. Wilcox, L.E. Bilston, D.C. Barton, R.M. Hall, Mathematical model for the viscoelastic properties of dura mater, in: J Orthop Sci, 2003: pp. 432–434.

[21] A. Bashkatov N., E. Genina A., Y. Sinichkin P., V. Kochubey I., N. Lakodina A., V. Tuchin V., Glucose and Mannitol Diffusion in Human Dura Mater, Biophys J 85 (2003) 3310–3318.

[22] J.T. Maikos, R.A.I. Elias, D.I. Shreiber, Mechanical properties of dura mater from the rat brain and spinal cord, J Neurotrauma 25 (2008) 38–51. https://doi.org/10.1089/neu.2007.0348.

[23] C. Persson, S. Evans, R. Marsh, J.L. Summers, R.M. Hall, Poisson's ratio and strain rate dependency of the constitutive behavior of spinal dura mater, Ann Biomed Eng 38 (2010) 975–983. https://doi.org/10.1007/s10439-010-9924-6.

[24] E. Mazgajczyk, K. Scigala, M. Czy, W. Jarmundowicz, R. BedZinski, Mechanical Properties of Cervical Dura Mater, Acta of Bioengineering and Biomaterials 14 (2012) 51–58.

[25] S. Shetye S., M. Deault M., C. Puttlitz M., Biaxial response of ovine spinal cord dura mater, J Mech Behav Biomed Mater 34 (2014) 146–153. 10.1016/j.jmbbm.2014.02.014 (accessed July 7, 2024).

[26] N. Ramo, S.S. Shetye, C.M. Puttlitz, Damage Accumulation Modeling and Rate Dependency of Spinal Dura Mater, J Eng Sci Med Diagn Ther 1 (2018). https://doi.org/10.1115/1.4038261.

[27]    D.R. Walsh, A.M. Ross, S. Malijauskaite, B.D. Flanagan, D.T. Newport, K.D. McGourty, J.J.E. Mulvihill, Regional mechanical and biochemical properties of the porcine cortical meninges, Acta Biomater 80 (2018) 237–246. https://doi.org/10.1016/j.actbio.2018.09.004.

[28]    D. De Kegel, J. Vastmans, H. Fehervary, B. Depreitere, J. Vander Sloten, N. Famaey, Biomechanical characterization of human dura mater, J Mech Behav Biomed Mater 79 (2018) 122–134. https://doi.org/10.1016/j.jmbbm.2017.12.023.

[29]    H.E. Aydın, C. Kızmazoglu, I. Kaya, B. Husemoglu, G. Sozer, H. Havıtcıoglu, A. Arslantas, Biomechanical properties of the cranial dura mater with puncture defects: An In Vitro study, J Korean Neurosurg Soc 62 (2019) 382–388. https://doi.org/10.3340/jkns.2018.0130.

[30]    J. Zwirner, M. Scholze, J.N. Waddell, B. Ondruschka, N. Hammer, Mechanical Properties of Human Dura Mater in Tension – An Analysis at an Age Range of 2 to 94 Years, Sci Rep 9 (2019). https://doi.org/10.1038/s41598-019-52836-9.

[31]    C. Yang, X. Yang, X. Lan, H. Zhang, M. Wang, Y. Zhang, Y. Xu, P. Zhen, Structure and mechanical characteristics of spinal dura mater in different segments of sheep's spine, Chinese Journal of Reparative and Reconstructive Surgery 33 (2019) 232–238. https://doi.org/10.7507/1002-1892.201807085.

[32]    J. Zwirner, B. Ondruschka, M. Scholze, G. Schulze-Tanzil, N. Hammer, Mechanical and morphological description of human acellular dura mater as a scaffold for surgical reconstruction, J Mech Behav Biomed Mater 96 (2019) 38–44. https://doi.org/10.1016/j.jmbbm.2019.04.035.

[33]    A.I. Lipovka, A. V. Dubovoy, D. V. Parshin, The study of the strength properties of the human dura mater: The experience of one research center, in: J Phys Conf Ser, IOP Publishing Ltd, 2020. https://doi.org/10.1088/1742-6596/1666/1/012062.

[34]    B. Pierrat, L. Carroll, F. Merle, D.B. MacManus, R. Gaul, C. Lally, M.D. Gilchrist, A. Ní Annaidh, Mechanical Characterization and Modeling of the Porcine Cerebral Meninges, Front Bioeng Biotechnol 8 (2020). https://doi.org/10.3389/fbioe.2020.00801.

[35]    A. Tamura, W. Yano, D. Yoshimura, S. Nishikawa, Mechanical Characterization of Spinal Dura Mater Using a PD-Controlled Biaxial Tensile Tester, J Mech Med Biol 20 (2020). https://doi.org/10.1142/S0219519420500232.

[36]    D.R. Walsh, A.M. Ross, D.T. Newport, Z. Zhou, J. Kearns, C. Fearon, J. Lorigan, J.J.E. Mulvihill, Mechanical characterisation of the human dura mater, falx cerebri and superior sagittal sinus, Acta Biomater 134 (2021) 388–400. https://doi.org/10.1016/j.actbio.2021.07.043.

[37]    Y. Su, Z. Li, H. Zhu, J. He, B. Wei, D. Li, Electrohydrodynamic Fabrication of Triple-layered Polycaprolactone Dura Mater Substitute with Antibacterial and Enhanced Osteogenic Capability, Chinese Journal of Mechanical Engineering: Additive Manufacturing Frontiers 1 (2022) 100026. https://doi.org/10.1016/j.cjmeam.2022.100026.

[38]    Q. Pearcy, M. Jeejo, M. Scholze, J. Tomlinson, J. Dressler, M. Zhang, J. Zwirner, Biomechanics of vascular areas of the human cranial dura mater, J Mech Behav Biomed Mater 125 (2022). https://doi.org/10.1016/j.jmbbm.2021.104866.

[39]    S. Çavdar, S. Sürücü, M. Özkan, B. Köse, A.N. Malik, E. Aydoğmuş, Ö. Tanış, İ. Lazoğlu, Comparison of the Morphologic and Mechanical Features of Human Cranial Dura and Other Graft Materials Used for Duraplasty, World Neurosurg 159 (2022) e199–e207.

[40] S. Cavelier, R.D. Quarrington, C.F. Jones, Mechanical properties of porcine spinal dura mater and pericranium, J Mech Behav Biomed Mater 126 (2022). https://doi.org/10.1016/j.jmbbm.2021.105056.

[41] M. Evin, P. Sudres, P. Weber, Y. Godio-Raboutet, P.-J. Arnoux, E. Wagnac, Y. Petit, Y. Tillier, Experimental Bi-axial tensile tests of spinal meningeal tissues and constitutive models comparison, Acta Biomater 140 (2022) 446–456.

[42] A. Tamura, S. Nishikawa, Effect of Anatomical Sites on the Mechanical Properties of Spinal Dura Subjected to Biaxial Stretching, J Eng Sci Med Diagn Ther 5 (2022). https://doi.org/10.1115/1.4053341.

[43] J.A. Niestrawska, M. Rodewald, C. Schultz, E. Quansah, T. Meyer-Zedler, M. Schmitt, J. Popp, I. Tomasec, B. Ondruschka, N. Hammer, Morpho-mechanical mapping of human dura mater microstructure, Acta Biomater 170 (2023) 86–96. https://doi.org/10.1016/j.actbio.2023.08.024.

[44] A. Sharma, J. Liao, L.N. Williams, Structure and mechanics of native and decellularized porcine cranial dura mater, Engineered Regeneration 4 (2023) 205–213. https://doi.org/10.1016/j.engreg.2023.02.004.

[45] A. Sharma, E. Moore, L.N. Williams, An in vitro study of micromechanics, cellular proliferation and viability on both decellularized porcine dura grafts and native porcine dura grafts, Biomedical Engineering Advances 6 (2023) 100108. https://doi.org/10.1016/j.bea.2023.100108.

[46] J. Zwirner, B. Ondruschka, M. Scholze, A. Thambyah, J. Workman, N. Hammer, J.A. Niestrawska, Dynamic load response of human dura mater at different velocities, J Mech Behav Biomed Mater 138 (2023). https://doi.org/10.1016/j.jmbbm.2022.105617.

[47] A. Tamura, S. Nishikawa, Dural Mechanical Responses to Load-Controlled Asymmetric Biaxial Stretch, J Mech Med Biol 23 (2023). https://doi.org/10.1142/S021951942350077X.

[48] A. Tamura, C. Sakaue, Effects of surface profile on porcine dural mechanical properties, Clinical Biomechanics (2024).

[49] X. Jin, J. Lee B., L. Leung Yee, L. Zhang, K. Yang H., A. King I., Biomechanical response of the Bovine Pia-Arachnoid Complex to Tensile Loading at Varying Strain-Rates, Stapp Car Crash J 50 (2006) 637–649. https://doi.org/https://doi.org/10.4271/2006-22-0025.

[50] X. Jin, C. Ma, L. Zhang, K.H. Yang, A.I. King, G. Dong, J. Zhang, Biomechanical Response of the Bovine Pia-Arachnoid Complex to Normal Traction Loading at Varying Strain Rates, 2007.

[51] X. Jin, K.H. Yang, A.I. King, Mechanical properties of bovine pia-arachnoid complex in shear, J Biomech 44 (2011) 467–474. https://doi.org/10.1016/j.jbiomech.2010.09.035.

[52] X. Jin, H. Mao, K.H. Yang, A.I. King, Constitutive modeling of pia-arachnoid complex, Ann Biomed Eng 42 (2014) 812–821. https://doi.org/10.1007/s10439-013-0948-6.

[53] N.L. Ramo, K.L. Troyer, C.M. Puttlitz, Viscoelasticity of spinal cord and meningeal tissues, Acta Biomater 75 (2018) 253–262. https://doi.org/10.1016/j.actbio.2018.05.045.

[54] G. Fabris, Z. M. Suar, M. Kurt, Micromechanical heterogeneity of the rat pia-arachnoid complex, Acta Biomater 100 (2019) 29–37. https://doi.org/10.1016/j.actbio.2019.09.044.

[55] N. Benko, E. Luke, Y. Alsanea, B. Coats, Mechanical characterization of the human pia-arachnoid complex, J Mech Behav Biomed Mater 120 (2021). https://doi.org/10.1016/j.jmbbm.2021.104579.

[56]  L. Qian, S. Wang, S. Zhou, Y. Sun, H. Zhao, Influence of pia-arachnoid complex on the indentation response of porcine brain at different length scales, J Mech Behav Biomed Mater 127 (2022). https://doi.org/10.1016/j.jmbbm.2021.104925.

[57]  A.R. Tunturi, Elasticity of the spinal cord, pia, and dentieulate ligament in the dog, J Neurosurg 48 (1978) 975–979.

[58]  H. Ozawa, T. Matsumoto, T. Ohashi, M. Sato, S. Kokubun, Mechanical properties and function of the spinal pia mater, 2004.

[59]  P. Aimedieu, R. Grebe, Tensile strength of cranial pia mater: preliminary results, J Neurosurg 100 (2004) 111–114.

[60]  H. Kimpara, Y. Nakahira, M. Iwamoto, K. Miki, Investigation of Anteroposterior Head-Neck Responses during Severe Frontal Impacts Using a Brain-Spinal Cord Complex FE Model, Stapp Car Crash J 50 (2006) 509–544.

[61]  Y. Li, W. Zhang, Y.C. Lu, C.W. Wu, Hyper-viscoelastic mechanical behavior of cranial pia mater in tension, Clinical Biomechanics 80 (2020). https://doi.org/10.1016/j.clinbiomech.2020.105108.

[62]  P. Sudres, M. Evin, E. Wagnac, N. Bailly, L. Diotalevi, A. Melot, P.-J. Arnoux, Y. Petit, Tensile mechanical properties of the cervical, thoracic and lumbar porcine spinal meninges, J Mech Behav Biomed Mater 115 (2021). https://doi.org/https://doi.org/10.1016/j.jmbbm.2020.104280.